\documentclass[12pt]{article}

\usepackage{newtxtext,newtxmath}

\usepackage{graphicx}

\usepackage[letterpaper,margin=1in]{geometry}

\renewenvironment{abstract}
	{\quotation}
	{\endquotation}

\date{}

\makeatletter
\renewcommand{\fnum@figure}{\textbf{Figure \thefigure}}
\renewcommand{\fnum@table}{\textbf{Table \thetable}}
\makeatother

\usepackage{scicite}

\usepackage{url}
\usepackage{lineno}
\modulolinenumbers[5]

\def\scititle{
	Tipping cascades may lock the Earth system on a pathway towards a melthouse
}
\title{\bfseries \boldmath \scititle}

\author{
	Victor~Couplet$^{1,2\ast}$,
	Wim~Thiery$^{1}$,
	Michel~Crucifix$^{2}$\and
	\small$^{1}$Department of Water and Climate, Vrije Universiteit Brussels, Brussels \& 1050, Belgium.\and
	\small$^{2}$Earth and Life Institute, UCLouvain, Louvain-la-Neuve \& 1348, Belgium.\and
	\small$^\ast$Corresponding author. Email: victor.couplet@vub.be\and
}

\begin{document} 

\maketitle



\begin{abstract} \bfseries \boldmath
The potential for the Earth system to transition irreversibly into a “hothouse Earth” has attracted considerable scientific and media attention but remains insufficiently explored in modeling studies. Using a reduced-complexity Earth system model that includes a dynamic carbon cycle and interacting tipping elements, we find no planetary threshold beyond which the Earth system irreversibly shifts into a hothouse state. Instead, tipping cascades can drive the system towards an irreversible “melthouse” state with over 10 m of sea-level rise. Such cascades may occur even under low-emission scenarios, preventing long-term stabilization of sea level at intermediate values. Our results overall highlight the substantial risk of a melthouse rather than a hothouse Earth, and call for drastic emission reductions to avert dramatic impacts on coastal cities and communities.
\end{abstract}

\noindent

The Intergovernmental Panel on Climate Change (IPCC) provides ample evidence that every degree of warming increases the risk and severity of disruptive effects on ecosystems and societies \cite{ipcc_climate_2021,intergovernmental_panel_on_climate_change_ipcc_global_2023}. Yet greenhouse gas emissions continue to rise, pushing global temperatures ever closer to breaching the 1.5°C Paris target and raising the possibility of crossing tipping points —thresholds beyond which critical Earth system components (tipping elements) may undergo self-sustaining and potentially irreversible changes \cite{lenton_tipping_2008,ipcc_climate_2021,armstrong_mckay_exceeding_2022,wang_mechanisms_2023,lenton_global_2023}.   

Tipping elements are interconnected and can amplify warming by altering planetary albedo (e.g., ice sheet collapse) or releasing greenhouse gases (e.g., Amazon rainforest dieback, permafrost thaw) \cite{wunderling_climate_2024,lenton_global_2023}. This suggests tipping one element could trigger transitions in others that would not tip independently, potentially leading to substantial warming even under low- to medium-emissions. Such a cascade would eliminate middle-of-the-road climate trajectories, leaving humanity with a binary choice: either drastically reduce emissions and stabilize the Earth system in a relatively cool state, or risk crossing a threshold beyond which a “hothouse Earth” state becomes inevitable \cite{steffen_trajectories_2018}.

While perspective pieces have drawn attention to these risks and sparked considerable scientific and media interest \cite{steffen_trajectories_2018, lenton_climate_2019, kemp_climate_2022}, this hypothesis largely lacks support from detailed model simulations, which are essential for more robust, evidence-based conclusions. Here we use the reduced-complexity climate model SURFER \cite{martinez_montero_surfer_2022,couplet_surfer_2025,couplet_tipping_2025} to investigate the possibility of a hothouse Earth scenario. 

\subsection*{Modeling approach}

Our model includes a dynamic carbon cycle and parameterizations for nine non-linear elements: six tipping elements (the Greenland and West Antarctic Ice Sheets, the East Antarctic Subglacial Basins, the rest of the East Antarctic Ice Sheet, the Amazon rainforest and the Atlantic Meridional Overturning Circulation (AMOC)) and three threshold-free non-linear elements (mountain glaciers, Arctic sea-ice, and the boreal permafrost). To explore possible future Earth system trajectories, we run the model until 100 000 CE across a set of 100 scenarios with increasing cumulative CO$_2$ emissions, under two configurations: a decoupled setup, in which interactions between non-linear elements and their feedbacks on global mean surface temperature (GMST) are disabled, and a coupled setup, in which these interactions and feedbacks are enabled (see Methods). 

The emissions scenarios are defined by the total amount of carbon released into the atmosphere after 2024 CE, ranging from 50~PgC to 5000~PgC by steps of 50~PgC (Fig. \ref{fig:example}A). We group them into three categories, based on the warming simulated in the decoupled model setup, which is independent of the state of the tipping and threshold-free elements. Low-emission scenarios (50–300~PgC) remain below 2°C of global warming. Medium-emission scenarios (350–1200~PgC) exceed 2°C but keep warming in 2100 below 2.7°C, the level projected under current policies (as of October 2025). Finally, scenarios with 1250-5000~PgC of additional emissions are classified as high-emission. 

\subsection*{Simulations with an illustrative parameter set}
\begin{figure} 
	\centering
	\includegraphics[width=0.9\textwidth]{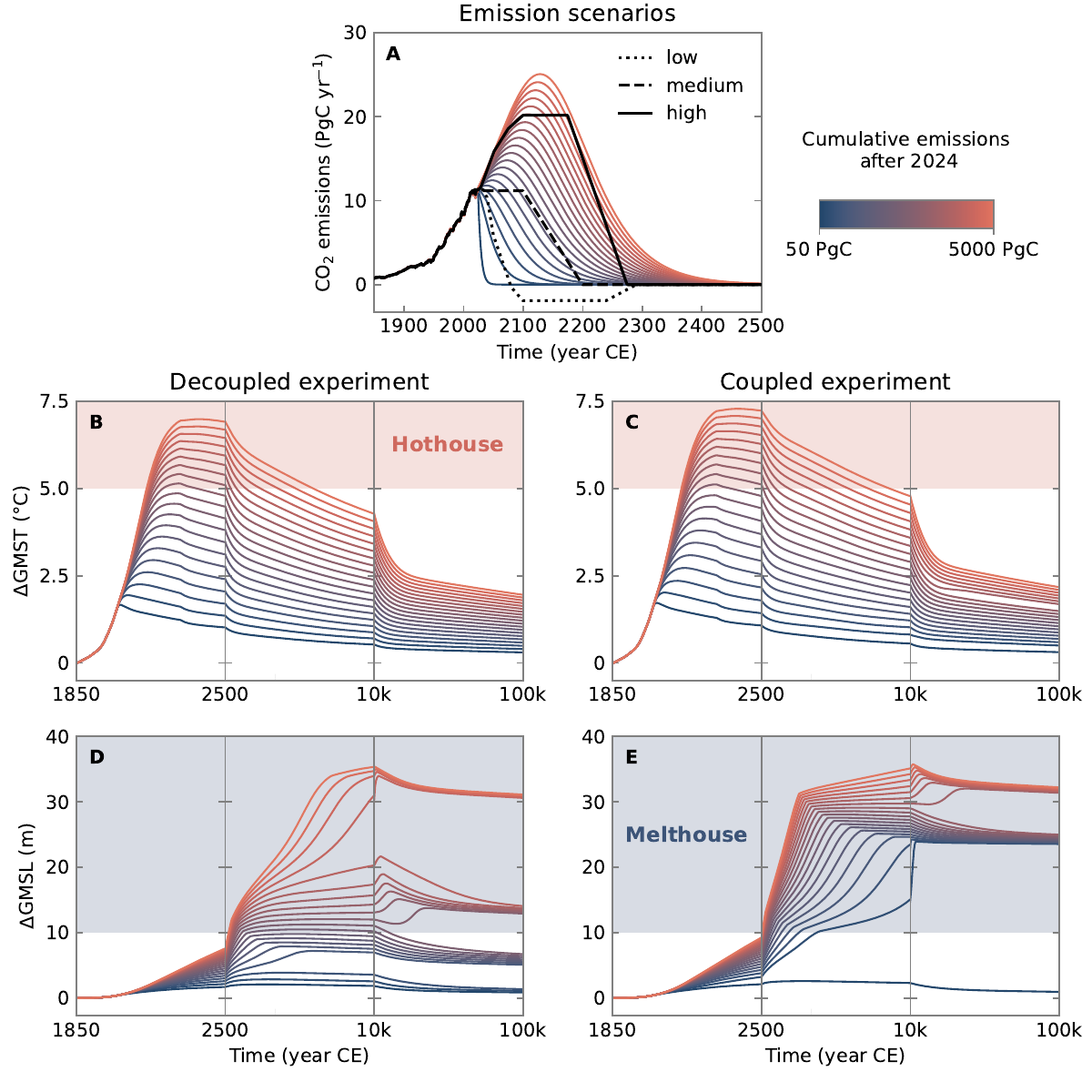} 
	\caption{\textbf{Experimental setup and illustrative outputs from SURFER}. \textbf{(A)} Emissions scenarios compared with the 'low', 'medium', and 'high' emission scenarios from the Scenario Model Intercomparison Project for CMIP7 \cite{van_vuuren_scenario_2025}. \textbf{(B-E)} Example outputs of global mean surface temperature (GMST) and global mean sea level (GMSL) for one illustrative parameter set for the climate tipping elements (tables S1–S2) under 20 of the 100 emission scenarios. Panels (B, D) show the decoupled experiment, where interactions among nonlinear elements and their feedbacks on temperature are disabled. Panels (C, E) show the coupled experiment, where these interactions and feedbacks are active. GMST and GMSL are expressed as anomalies relative to 1850 CE, the start of the simulations.}
	\label{fig:example}
\end{figure}

We begin by examining model simulations with a specific, illustrative set of model parameters (Fig. \ref{fig:example}B-E, tables S1 and S2). In the decoupled runs, the AMOC, West Antarctic Ice Sheet, Greenland Ice Sheet, East Antarctic Subglacial Basins, and Amazon rainforest first collapse under the 250, 600, 2550, 3850, and 4000~PgC emission scenarios, respectively. In the coupled runs, interactions between the elements and their feedback on GMST trigger a cascade where the AMOC, the West Antarctic Ice Sheet, and the East Antarctic Subglacial Basins all tip together under the 100~PgC scenario, while the AMOC’s stabilizing influence delays the tipping of the Greenland Ice Sheet and Amazon rainforest to the 3300 and 4950~PgC scenarios, respectively. Comparing the two experimental setups shows that the tipping elements have a negligible impact on global mean temperature (Fig. \ref{fig:example}B-C). We define a hothouse state as a situation in which Earth's globally averaged surface temperature increases by 5°C or more relative to the pre-industrial period, and regardless of whether interactions between the tipping elements and their feedback on GMST are included, such a hothouse is reached only under the highest emission scenarios (Fig. \ref{fig:example}B-C). In contrast, tipping elements have a pronounced impact on global mean sea level (GMSL) trajectories and can cause substantial jumps in GMSL between scenarios (Fig. \ref{fig:example}D-E). Notably, low-emission scenarios can lead to a GSML rise of at least 10 m above the pre-industrial level. We refer to this situation as a melthouse state. 

Substantial uncertainties surround the critical threshold of tipping elements and the time scales over which they might collapse \cite{armstrong_mckay_exceeding_2022,lenton_global_2023}. Arguably, even larger uncertainties surround interactions between elements \cite{wunderling_climate_2024}. To account for this, we sample 1,000 different parameter sets spanning uniformly values reported in the literature \cite{armstrong_mckay_exceeding_2022,lenton_global_2023,wunderling_climate_2024,rahmstorf_thermohaline_2005} (see Methods, and tables S3-S4). We run SURFER using each of these parameter sets, and analyze a total of $100 \times 1,000 = 100,000$ model runs in the decoupled setup (the \textit{decoupled experiment}) and  $100,000$ runs in the coupled setup (the \textit{coupled experiment}).

\subsection*{Influence of fully coupled tipping elements on the modeled climate response}\label{sec2}

\begin{figure}
    \centering
    \includegraphics[width=0.6\textwidth]{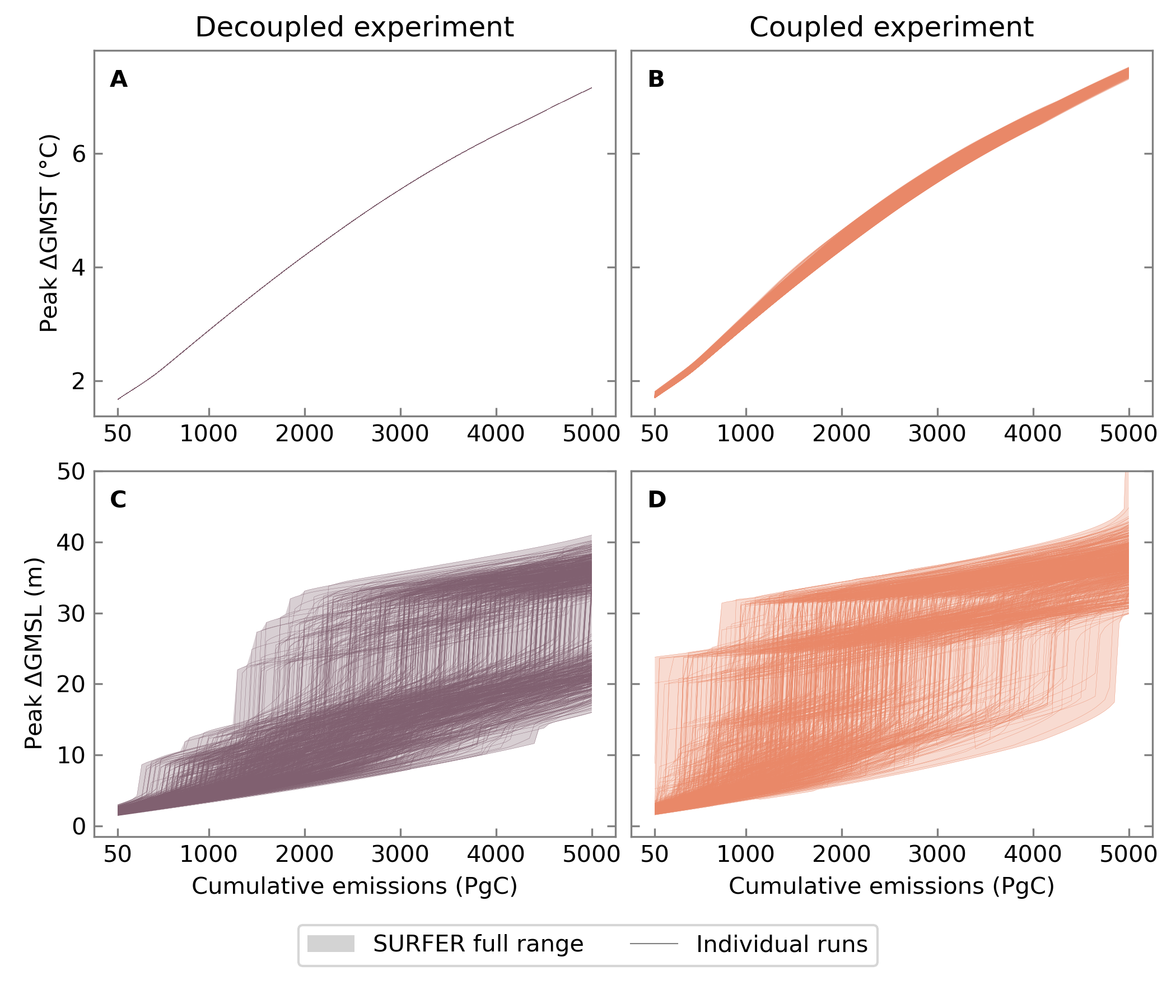}
    \caption{\textbf{Fully coupling tipping and other nonlinear elements in SURFER moderately affects global temperature but substantially impacts sea level rise.} \textbf{(A, B)} Peak global mean surface temperature (GMST) and \textbf{(C, D)} peak global mean sea level (GMSL) as functions of cumulative CO$_2$ emissions for the decoupled and coupled experiments, expressed relative to 1850 CE. Shaded areas indicate the full range across all simulations, and thin solid lines show individual runs. In the decoupled experiment, GMST is independent of the states of tipping and threshold-free elements, producing indistinguishable results across the 1,000 parameter sets in panel (A). The timing of peak values may differ between the decoupled and coupled experiments for a given parameter set and emission scenario.}
    \label{fig:elements_impact}
\end{figure}

We compare the coupled and decoupled experiments to assess how interactions among non-linear elements and their feedbacks on GMST modify the climate response. In the coupled experiment, including these feedbacks produces slightly higher temperatures than in the decoupled setup (Fig. \ref{fig:elements_impact}A-B). The difference generally grows with cumulative emissions and is the largest at intermediate time scales, reaching 0.37-0.73°C at time $t=5000$ year CE under 5000~PgC of emissions (fig. \ref{fig:impacts_times_T}). By this stage, natural carbon sinks have begun to reduce temperatures, so the increase in peak warming due to tipping and other non-linear threshold-free elements never exceeds 0.46°C. Our results suggest that, at any given time, global mean temperature scales nearly linearly with cumulative emissions. This holds both on the century time scale, consistent with IPCC assessments \cite{ipcc_climate_2021}, and on longer time scales (fig. \ref{fig:impacts_times_T}). Overall, the influence of tipping and other non-linear elements is small compared to the choice of emission scenario, in line with recent studies \cite{wang_mechanisms_2023,deutloff_high_2025} which modeled the influence of various tipping elements on temperature trajectories up to 2300 and 2500 using the reduced-complexity climate model FaIR \cite{smith_fair_2018}.


For a given emissions scenario, GMSL is generally higher in the coupled experiment than in the decoupled experiment (Fig. \ref{fig:elements_impact}C-D). Strikingly, in the coupled setup, GMSL can rise by up to 20 m even under low-emissions scenarios ($\leq$300~PgC, peak warming below $\sim$2°C). These elevated responses are driven by interactions between tipping elements that raise the risk of collapse of both the West Antarctic Ice Sheet and the East Antarctic Subglacial Basins \cite{couplet_tipping_2025}. In some model simulations, however, sea-level rise is lower in the coupled setup due to the stabilizing influence of the Atlantic Meridional Overturning Circulation (AMOC) on the Greenland Ice Sheet. Overall, our results indicate a non-linear long-term response of GMSL to cumulative emissions irrespective of the parameter set, with potentially significant differences even between similar emission scenarios (Fig. \ref{fig:elements_impact}C-D, fig. \ref{fig:impacts_times_slr}).

\subsection*{Hothouse and melthouse risk}

\begin{figure}
\centering
    \includegraphics[width=0.6\textwidth]{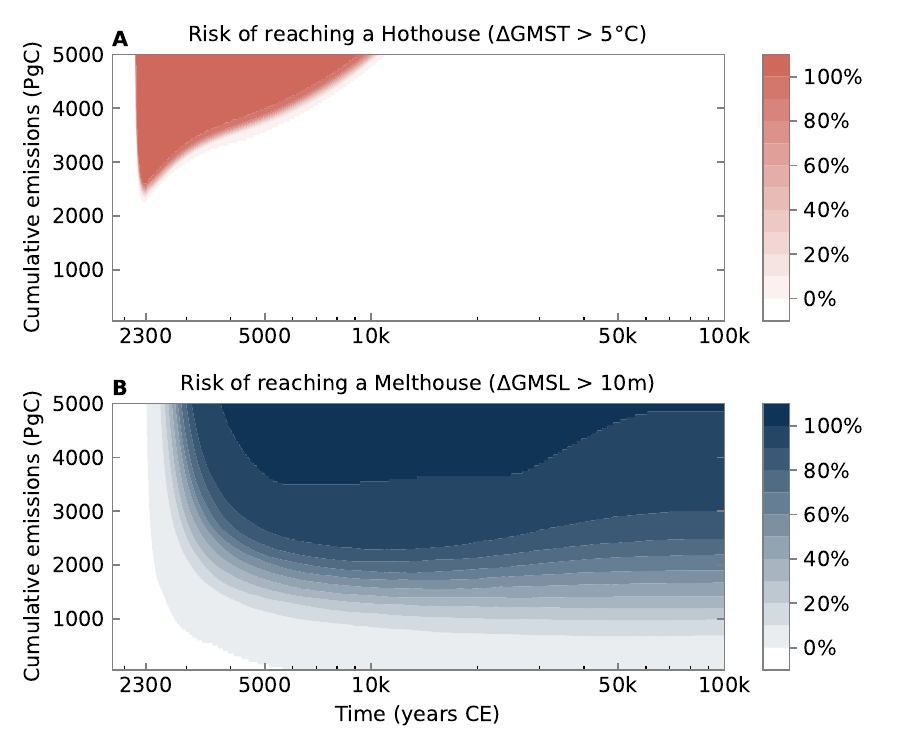}
    \caption{\textbf{A hothouse state occurs only under high-emissions scenarios, whereas a melthouse state can be reached under low-emissions}. \textbf{(A)} Risk of reaching hothouse and \textbf{(B)} risk of reaching a melthouse state as functions of time and cumulative emissions in the coupled experiment. For each time $t$ and emission scenario, the risk is defined as the percentage of ensemble simulations which exceed the reference thresholds : $\Delta \text{GMST} > 5$°C for the hothouse, and $\Delta \text{GMSL} > 10$ m  for the melthouse. Parameters are sampled from uniform distributions that span reported ranges. Risk values between 0\% and 100\% are here defined as the frequencies of occurrence of hothouse or melthouse events in our experimental design}.
    \label{fig:risk}
\end{figure}

We then investigate the hothouse or melthouse risk for a given emission scenario and time, with risk defined as the percentage of ensemble simulations (across all parameter sets) in which GMST exceeds 5 °C or GMSL exceeds 10 m, both relative to the pre-industrial baseline (1850 CE). In the coupled experiment and with a mid-range climate sensitivity of 3.5 °C, hothouse risk is non-zero only under high-emission scenarios ($\geq$2300~PgC; Fig. \ref{fig:risk}A). Even then, the state is reversible after several thousand years due to natural carbon sinks. The transition between 0 and 100\% risk is sharp, indicating that parameter uncertainty has little influence on hothouse risk and that the state is determined primarily by cumulative emissions rather than tipping dynamics; consistent with Figure \ref{fig:elements_impact}B.

In contrast, we find a non-zero melthouse risk under low-emission scenarios (Fig. \ref{fig:risk}B), with a melthouse state possible by $\sim$4000-5000 CE for scenarios with less than 300~PgC of emissions (peak warming below $\sim$2°C). The risk then rises gradually with cumulative emissions, reflecting dependence on the poorly constrained sensitivities of the ice sheets. For the highest emission scenarios ($\geq$ 3550~PgC, peak warming above $\sim$6°C), all parameter sets lead to a melthouse state, which could occur as early as the year 2314 CE. This is consistent with IPCC projections, which do not rule out up to 15 meters of sea-level rise by 2300 under SSP5-8.5 \cite{ipcc_climate_2021}. Once established, a melthouse state typically persists for tens of thousands of years, owing to the slow pace of natural cooling and the inherently slow recovery rate of ice sheets.

\subsection*{Tipping cascades and diverging Earth system trajectories}

Can small emission increments trigger step changes in global warming and sea-level rise? We assess this by comparing model runs across successive scenarios in both the coupled and decoupled experiments, using the responses of tipping elements to identify tipping events and cascades. In each simulation, the six tipping elements are assigned a status which is either \textit{tipped} or \textit{not tipped} (see Methods). For a given experiment and parameter set, a \textit{tipping event} is said to occur in scenario $j$ if at least one element changes status relative to scenario $j-1$, which corresponds to 50~PgC less anthropogenic CO$_2$ emissions. Events involving a single element are classified as \textit{single tipping events}, while those with $n>1$ elements are classified as \textit{tipping cascades} of size $n$. To assess their climatic response, we compare global temperature and mean sea-level rise between scenarios $j$ and $j-1$, testing whether each event drives divergent trajectories and commits the Earth system to a hothouse or melthouse state.

\begin{figure}
    \centering
    \includegraphics[width=0.6\textwidth]{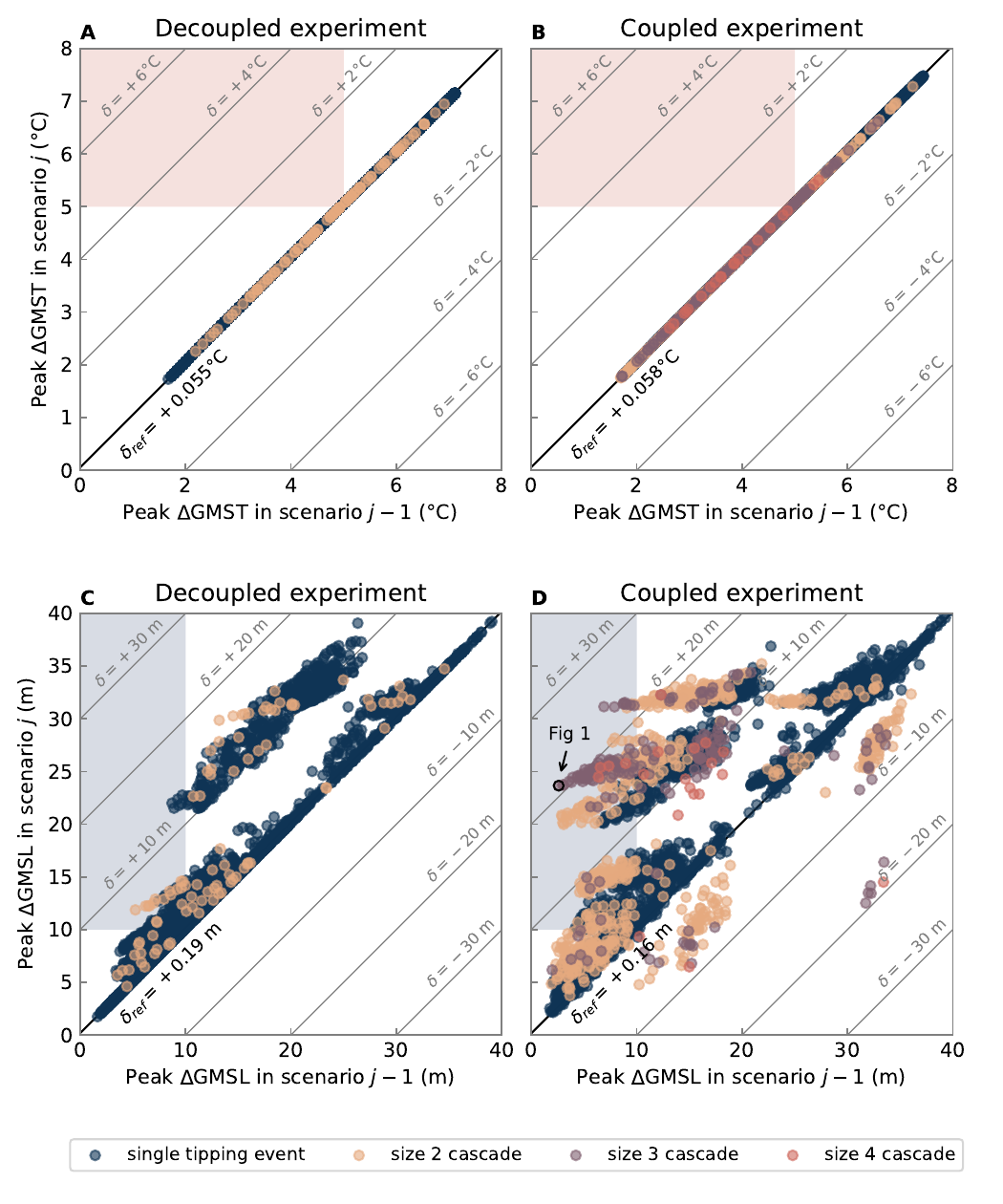}
    \caption{\textbf{Tipping cascades can drive large increases in sea level rise}. Influence of tipping events on peak warming \textbf{(A, B)} and peak global mean sea level \textbf{(C, D)}. Each observed tipping event is shown as a point, colored by event type. The climate impact of each event is assessed by comparing the coordinates of the corresponding point, which represent the peak warming or sea level rise in scenario $j$ (where the event occurred) and in scenario $j-1$ (50~PgC lower cumulative emissions). The thick black diagonal in each plot has equation $y=x+\delta_{ref}$, where $\delta_{ref}$ is the mean difference in peak warming or sea level rise between adjacent scenarios without tipping events. Points on this line have a negligible influence, whereas deviations from it indicate the strength of the tipping event’s effect. The thin gray diagonals indicate contours of equal climate response (GMST or GMSL) and are provided as visual guides. Shaded regions highlight tipping events that push peak warming past 5°C (A, B) or sea level rise past 10~m (C, D) above pre-industrial levels (1850~CE), starting from below these thresholds. In the decoupled experiment (A,C), the interactions between elements are disabled, so events identified as size-2 cascades arise only from coincident tipping of independent elements.}
    \label{fig:gaps}
\end{figure}

Our results show that tipping events have a limited influence on peak warming, with a maximum increase of $\sim$0.09°C between successive scenarios in the coupled experiment (Fig. \ref{fig:gaps}A-B). Two main factors explain this. First, most tipping elements in our model, including the Greenland Ice Sheet, the West Antarctic Ice Sheet, the East Antarctic Subglacial Basins, and the Amazon rainforest, contribute little to global radiative forcing. The East Antarctic Ice Sheet has the potential to contribute up to $\sim$0.7°C of warming in our model, but only melts completely once under the prescribed scenarios. Permafrost can release large amounts of CO$_2$ and CH$_4$ into the atmosphere, yet it responds mostly linearly to temperature and is not directly coupled to other tipping elements, limiting the potential for substantial changes in thaw emissions between successive scenarios. Second, as outlined in the previous section, most elements evolve on slower timescales than anthropogenic forcing, so their impact on temperature becomes pronounced only after peak warming, which is primarily determined by anthropogenic emissions, and occurs between 2100 and 2500 CE for the scenarios considered here. Even on longer time scales, the GMST difference between two successive scenarios never exceeds 0.59°C in the coupled experiment, or 0.29°C when excluding the unique run with complete East Antarctic Ice sheet meltdown (fig. \ref{fig:gaps_times_T}). Together, these results suggest tipping cascades are unlikely to drive the Earth system into a hothouse state in a way that would prevent the stabilization of global temperatures at intermediate levels. 

The outlook is very different for sea level rise. In both the decoupled and coupled experiments, the tipping of a single element can drive substantial increases in peak GMSL between successive scenarios: around 5 m for the collapse of the Greenland Ice Sheet or the West Antarctic Ice Sheet, and up to 10 m for the East Antarctic Subglacial Basins (Fig. \ref{fig:gaps}C-D). Because collapse typically occurs after partial melting, the GMSL contributions from tipping events are generally smaller than the full sea-level rise potential of the affected ice sheets.

Including interactions between elements in the coupled experiment increases the occurrence of cascades, which can drive substantial sea-level rise when they involve multiple cryospheric elements (Fig. \ref{fig:gaps}C-D). Overall, 1.1\% of parameter sets generate coupled runs in which single tipping events or cascades occur under low emission scenarios (50-300~PgC) and result in a peak GMSL rise exceeding 10~m, with an increase of at least 1 m compared to the scenario with 50~PgC less emissions. For medium-emission scenarios (350-1200~PgC) that fail to implement the Paris Agreement, this fraction rises to 24.2\%. Among these high-GMSL-rise tipping events, about one-third involve at least two elements that drive sea level rise, and the largest divergences in trajectories arise when the East Antarctic Subglacial Basins are involved. In such instances, a difference of only five years’ worth of emissions at today’s rate can trigger cascades that drive up to $\sim$20m of additional sea level, although these consequences manifest only after several thousand years (see figure \ref{fig:gaps_times_slr}, and the example in Figure \ref{fig:example}). This underlines the existence of a threshold beyond which tipping cascades may lock the Earth system on a pathway towards a melthouse, preventing the stabilisation of long-term GMSL rise at intermediate values. Where such a threshold lies depends on the behaviour of the ice sheets, and our results indicate that it could be reached under scenarios consistent with current policies, or even in scenarios that limit warming below 2°C, albeit with a reduced risk. 

\subsection*{Future trajectories and Earth system stability}

We synthesize our results and provide a visual overview of possible Earth system trajectories in Figure \ref{fig:distributions_3D}. It shows the time evolution of temperature and sea-level rise distributions across all model parameter sets and emission scenarios. The figure can be viewed as a model-derived counterpart to the conceptual stability landscape proposed by \cite{steffen_trajectories_2018} (reproduced in figure \ref{fig:stability_vs_density}). Here, high values (that is, troughs in the figure) indicate states that consistently occur in SURFER across multiple emissions scenarios and possible tipping element characteristics, and which may therefore be considered as likely stable (see supplementary text). 

The distributions for GMST (Fig. \ref{fig:distributions_3D}, A and C) show no clear divergence in Earth system trajectories, consistent with Figures \ref{fig:elements_impact}B and \ref{fig:gaps}B. This implies that a mostly continuous range of temperatures can be reached at any given time, with outcomes  determined primarily by anthropogenic emissions. In contrast, we observe clear divergences in GMSL trajectories, consistent with Figures \ref{fig:elements_impact}D and \ref{fig:gaps}D. On centennial time scales, the distribution of GMSL rise is unimodal (Fig. \ref{fig:distributions_3D}B), with a high-rise tail gradually emerging. After several thousand years, multiple modes appear, reflecting different combinations of tipped elements (Fig. \ref{fig:distributions_3D}D; fig. \ref{fig:distributions_2D}). A distribution presenting multiple modes is a robust feature, even though the precise number of modes may depend on parameter choices. We find it significant that some high-GMSL-rise modes are attainable even under low-emission scenarios, though they are more likely to be reached under high cumulative emissions. For instance, under a 300~PgC low-emission scenario, the risks of ending in the $\sim$15 m, $\sim$21 m, $\sim$25 m, and $\sim$32 m modes are 0.1\%, 0.5\%, 0.4\%, and 0\%, respectively. Under a 1200~PgC medium-emission scenario, these risks rise to 9.1\%, 4.8\%, 9.5\%, and 2.8\% (see table \ref{tab:scenario_modes}). This indicates that strong emissions cuts, limiting warming below 2°C, can substantially lower the risk of  committing to extreme long-term sea-level rise. 

\subsection*{Discussion}\label{sec12}

We have investigated the Earth system's response to anthropogenic emissions using the reduced-complexity model SURFER, which incorporates parameterizations for nine tipping and other non-linear elements. Our simulations show no evidence of a planetary threshold beyond which the Earth system irreversibly evolves into a hothouse state, as hypothesized by \cite{steffen_trajectories_2018}. Instead, global mean temperature responds approximately linearly to cumulative emissions at all time scales, indicating that future warming is primarily controlled by greenhouse gas emissions rather than tipping cascades. However, we emphasize that these findings do not eliminate the threat of severe climate impacts. 

First, climate impacts on natural and human systems do not start beyond the arbitrary 5°C benchmark we use to define a hothouse state. They are already being felt today and will intensify with every further increment of warming \cite{ipcc_climate_2021,intergovernmental_panel_on_climate_change_ipcc_global_2023}. Moreover, individual tipping elements may cause severe regional consequences even under relatively low levels of warming. Our model is not designed to capture these impacts.  

Second, our findings depend on the set of tipping elements and non-linear processes included in our model, and missing feedbacks could substantially amplify warming \cite{ripple_many_2023}. In SURFER v3.0, the overall carbon–climate feedback is relatively weak, so temperature changes exert only a modest influence on the carbon cycle \cite{couplet_surfer_2025}. This feedback is strengthened in the current version by including emissions from Amazon dieback and permafrost thaw, but additional mechanisms could amplify it further, for example temperature- and pH-sensitive ocean carbon pumps  \cite{henson_uncertain_2022,planchat_carbonate_2024,rothman_characteristic_2019}  or methane hydrate destabilization \cite{archer_ocean_2009,renssen_modeling_2004}. Methane hydrates store an estimated 1,000–2,000~PgC of methane in ice-like structures beneath the seabed, and  while most emissions would likely remain in sediments or oxidize before reaching the atmosphere, the residual long-term feedback could rival that from permafrost thaw \cite{minshull_mechanistic_2016,ruppel_interaction_2017}. Beyond the carbon cycle, cloud feedbacks remain a major source of uncertainty, with the potential to drive high-end equilibrium climate sensitivities of 5 °C or more \cite{bjordal_equilibrium_2020}. At CO$_2$ concentrations above~1,200 ppm, they could even trigger instabilities resulting in~8 °C of additional global surface warming \cite{schneider_possible_2019}. In SURFER, cloud feedbacks are included through the climate feedback parameter whose value is set to obtain an equilibrium climate sensitivity of 3.5°C, but higher sensitivities increase the risk of reaching a hothouse state (see supplementary text). An interplay between high climate sensitivity and strong carbon-climate feedbacks could further exacerbate the risk of observing large warming, even under low-emission scenarios \cite{kaufhold_interplay_2025}.

Finally, while we find no evidence of a hothouse threshold, our experiments show that tipping cascades could lock the Earth system on a melthouse trajectory, characterized by high sea level rise rather than high temperatures. Once triggered, such cascades would commit the Earth system to multi-meter sea-level rise, preventing stabilization at lower levels. This could occur under emission pathways consistent with current policies and possibly even if warming is kept below 2°C. Although much of the rise would unfold over millennia, high-emission scenarios could still lead to more than 10 m by $\sim$2300. This highlights the imperative of rapid, sustained emissions reductions in our lifetimes, not only to limit near-term warming, but also to avoid committing future generations to irreversible, multi-millennial sea level rise.

\subsection*{Methods}

We analyse the same model outputs as in \cite{couplet_tipping_2025}, and a detailed explanation of the model and the overall experimental setup can be found there. We summarise the key points below.

\subsubsection*{Tipping elements representation in SURFER}

SURFER is a reduced-complexity box model describing the exchanges of carbon and heat between the land, the atmosphere, and the ocean. We use version 3.0 of SURFER, which includes dynamic alkalinity cycling, CaCO$_3$ sediments, and weathering processes,  enabling simulations of the coupled carbon-climate system over multi-millennial timescales \cite{couplet_surfer_2025}. The model reproduces century-scale responses to emissions, including global mean temperature change, ocean pH decline, and dynamics of carbon sinks, consistent with current-generation Earth system models. It also produces long-term atmospheric CO$_2$ trajectories within the range of intermediate-complexity models \cite{couplet_surfer_2025}. 

We have extended the model by incorporating parametrisations for six tipping elements: the Greenland Ice Sheet (GRIS), the West and East Antarctic Ice Sheets (WAIS and EAIS), the East Antarctic Subglacial Basins (EASB), the AMOC, and the Amazon rainforest (AMAZ); as well as three threshold-free nonlinear elements: mountain glaciers, Arctic sea-ice (ASI), and boreal permafrost (PERM) \cite{couplet_tipping_2025}. The state evolution of these elements is modelled using the following equations :

\begin{equation}
\frac{\mathrm{d} x_i}{\mathrm{~d} t}\!=\!\left[\underbrace{\overbrace{-x_i^3+a_{2,i} x_i^2+a_{1,i} x_i +c_{0,i}}^{\text{internal dynamics}}+\,c_{1,i}\!\left(\delta T_\mathrm{U}+d\cdot\!\sum_j \epsilon_{ij} \delta L_j q_{*,i}\right)}_{H_i}\right]\!\mu\left(x_i, H_i\right), \label{eq:elements_4}
\medskip
\end{equation}
with $i\in \left\{\text{GRIS},\text{ WAIS},\text{ EASB},\text{ EAIS},\text{ ASI},\text{ AMOC},\text{ AMAZ},\text{ PERM}\right\}$. The adimensional variables $x_i$ represent the states of the elements and evolve between $0$ and $1$, a value of $1$ indicating the pre-industrial state and $0$ the collapsed state. The evolution of mountain glaciers is described by a  different parametrisation, which originates from previous versions of the model (see supplementary text).  

The parameters $(a_2,a_1,c_1,c_0)$ in Eq. \ref{eq:elements_4} are related to key quantities, such as the temperature thresholds $T_+$ for tipping elements, and the temperature of maximum sensitivity to forcing, $T_I$, for threshold-free nonlinear elements (see \cite{couplet_tipping_2025}). These temperatures are calibrated based on the review by \cite{armstrong_mckay_exceeding_2022}. The factor $\mu$ is given by 
\begin{equation}
\mu\left(x_i, H_i\right)= \begin{cases}1 / \tau_i^{+} & \text {if } H_i>0 \text {  and  } 0<x_i< 1 , \\
1 / \tau_i^{-} & \text {if } H_i<0 \text {  and  } 0<x_i< 1 , \\
0 & \text { if } x_i \leq 0 \text { or } x_i\geq 1,\end{cases}
\label{eq:tau}
\end{equation}
where $\tau_+$ and $\tau_-$ represent the time scales associated with growth and decline of the element. 
These are also calibrated based on \cite{armstrong_mckay_exceeding_2022}. Each element $i$ is forced by the global mean surface temperature anomaly $\delta T_U$ (relative to 1850 CE), and it can also be forced with any other quantity $\delta  L_j$ simulated by SURFER, allowing for numerous coupling possibilities between the different elements. In particular, the parameters $\epsilon_{ij}$ control the sign and strength of interactions. They are calibrated based on the literature and expert judgements synthesised in the recent review of tipping interactions by \cite{wunderling_climate_2024}. The coefficient $d$ allows one to scale all interactions simultaneously; it is set to $1$ in the coupled experiment and to $0$ in the decoupled experiment. The coefficient $q_{*,i}$ is given by 
\begin{equation}
    q_{*,i} = \begin{cases} T_{+,i} &\text{ for  } i \in \text{GRIS, WAIS, EASB, EAIS, AMOC, AMAZ,}
    \\
     T_{I,i}   &\text{ for  } i \in \text{AIS, PERM.}
    \end{cases}
\end{equation}
and is introduced purely for mathematical convenience. 

We include the impacts of the ice sheets (GRIS, WAIS, EASB, EAIS) on global mean temperatures through albedo changes based on \cite{wunderling_global_2020}. In our model, the maximum impact they can have on mean temperatures for an ECS of 3.5°C is 0.15°C, 0.06°C, 0.06°C, and 0.7°C, respectively. We also consider the additional greenhouse gas emissions from the Amazon rainforest dieback and boreal permafrost thaw. In the model, AMAZ can release a maximum of 75~PgC of CO$_2$, while permafrost can release a maximum of 300~PgC of CO$_2$ and 7.5~PgC of methane. Other feedbacks from the elements on temperature, such as the albedo changes resulting from Arctic sea ice loss or Amazon dieback, are considered to be already included in the climate feedback parameter $\beta$. This parameter is fixed in SURFER to provide an equilibrium climate sensitivity of 3.5°C. We neglect the influence of an AMOC collapse on global mean temperature and oceanic carbon release, as capturing these responses is beyond the capabilities of SURFER. However, the local effects of an eventual AMOC collapse on other elements (e.g., ASI, PERM, GRIS, AMAZ) are taken into account through the coupling terms in equation \ref{eq:elements_4}. GMSL rise is computed as the sum of six distinct contributions: oceanic expansion (thermosteric sea level rise), and meltwater from mountain glaciers, Greenland Ice Sheet, West Antarctic Ice Sheet, East Antarctic Subglacial Basins, and East Antarctic Ice Sheet. Sea-level projections from the model are consistent with IPCC assessments up to 2150 and fall within the range of multi-millennial estimates from previous studies (see supplementary text).

\subsubsection*{Dealing with uncertainties}

 To deal with the uncertainty surrounding the behaviour of tipping and other non-linear elements, we sample their critical temperatures ($T_+$ or $T_I$), their intrinsic dynamics time scales ($\tau_-$ and $\tau_+$), and the strength of their interactions (parameters $\epsilon_{ij}$) in uniform distributions (tables \ref{tab:elements_params_reduced}-\ref{tab:interaction_params}) using Latin hypercube sampling. We create an ensemble of $1,000$ different parameter sets and run the model for each of these parameter sets. We emphasize that, by sampling the parameters of tipping elements and their interactions from uniform distributions, our goal is to explore the range of possible Earth-system trajectories rather than to produce a strictly quantitative probabilistic projection of the future. This is why we use the term \textit{risk} rather than \textit{probability}. Specifically, we define the risk of an event as the percentage of ensemble simulations across all parameter sets in which the event occurs. For example, if a melthouse state arises in 13 out of 1000 simulations under the 300 PgC scenario, the melthouse risk for that scenario is 1.3\%. While this definition has the mathematical properties of a probability measure, we avoid calling it a probability because it has not been derived through a formal Bayesian treatment.

\subsubsection*{Emission scenarios}

We use emission scenarios that follow historical CO$_2$ emissions from 1850 to 2024 CE \cite{friedlingstein_global_2025}. After that,  between 50 and 5000~PgC of CO$_2$ are released following a logistic equation. Methane emissions are the same for all scenarios, following historical emissions from 1850 CE until 1989 CE \cite{jones_national_2023}, then following the optimistic SSP1-2.6 scenario \cite{riahi_shared_2017,gidden_global_2019}. We neglect the forcing from other greenhouse gases, ozone, and aerosols. We order scenarios by increasing cumulative emissions and index them such that scenario $j$ corresponds to  $j\cdot 50$~PgC of CO$_2$ emissions produced after 2024. Throughout this paper, we refer to scenarios by their CO$_2$ cumulative emissions after 2024 rather than their total cumulative emission value since 1850. The latter can be directly computed by adding 723~PgC to the post-2024 value. The classification of scenarios between low-emission (50-300~PgC), medium-emission (350-1200~PgC) and high-emission (1250-5000~PgC) is based on the nomenclature introduced in \cite{van_vuuren_scenario_2025}.

\subsubsection*{The coupled and decoupled experiments}

For each parameter set and emissions scenario, we run the model from 1850 CE to 100,000 CE. This constitutes a single experiment with $1,000 \times 100 = 100,000$ model runs. We perform several such experiments for different model settings. The two main experiments discussed in the paper are the \textit{decoupled} and \textit{coupled} experiments. In the decoupled experiment, interactions between the elements and their feedback on temperature are switched off, while in the coupled experiment, they are taken into account. In the supplementary material, we also test a higher climate sensitivity and stronger interactions between the non-linear elements.

\subsubsection*{Tipping cascades}

In every model run, we assign to the six tipping elements a \textit{tipping status}, which is either \textit{tipped} or \textit{not tipped}. We assign to an element $i$ the status \textit{tipped} if there exists a time $t$ ($1850<t<100000$) such that $x_i(t)<0.25$; otherwise, we assign the status \textit{not tipped}. This means that a component of the climate system is considered to have tipped if its state variable drops below 25\% of its initial normalized value at any point during the simulation, indicating a substantial and likely irreversible shift in its state. In a given experiment and for a given set of parameters, we say there is a \textit{tipping event} for scenario $j$ if at least one tipping element has changed status in scenario $j$ compared to scenario $j-1$. If only one element has changed status, we classify the event as a \textit{single tipping event}. If $n>1$ elements have changed status, we call it a \textit{tipping cascade} of size $n$. Under this definition, cascades may, in principle, include cases where elements tip independently rather than through interactions. In fact, we identify 94 cascades in the decoupled experiment, where tipping interactions are excluded. This implies that, out of the 904 cascades found in the coupled experiment, approximately 10.4\% ($94/904\times100$) include instances of independent tipping. This small fraction is not problematic, as our analysis focuses on the climate response associated with cascades, which depends only on the set of tipped elements rather than the specific causal chain of events.

\subsubsection*{Hothouse and melthouse}

The term \textit{hothouse} broadly refers to a planet several degrees warmer,  with wholly or partially melted ice sheets, and the word has been used previously with varying meanings. For instance, when \cite{steffen_trajectories_2018} refers to hothouse earth, they are talking about a state akin to the Mid-Pliocene epoch, where GMST was 4-5°C warmer than today, and sea levels 10-60 m higher. Authors that consider Earth's history on a larger time scale would still view it as a cool greenhouse [e.g. \cite{kidder_human-induced_2012}] or even a coolhouse [e.g. \cite{westerhold_astronomically_2020}], reserving the term hothouse for Eocene-like climates, with 10+°C temperatures and ice-free conditions. Here we adopt the \cite{steffen_trajectories_2018} framing, but consider GMST and GMSL separately. We define a hothouse as an Earth system state characterized by GMST exceeding 5°C above pre-industrial levels, and a melthouse as a state with GMSL more than 10 m above pre-industrial levels. Pre-industrial conditions are taken as those of 1850 CE, the baseline for our simulations. While the Global Tipping Points Report 2023 introduced the term 'wethouse' for states with tens of meters of sea-level rise, we use melthouse to avoid confusion with hydrological cycle changes, which are not represented in SURFER.

\begin{figure}
    \centering
    \includegraphics[width=\textwidth]{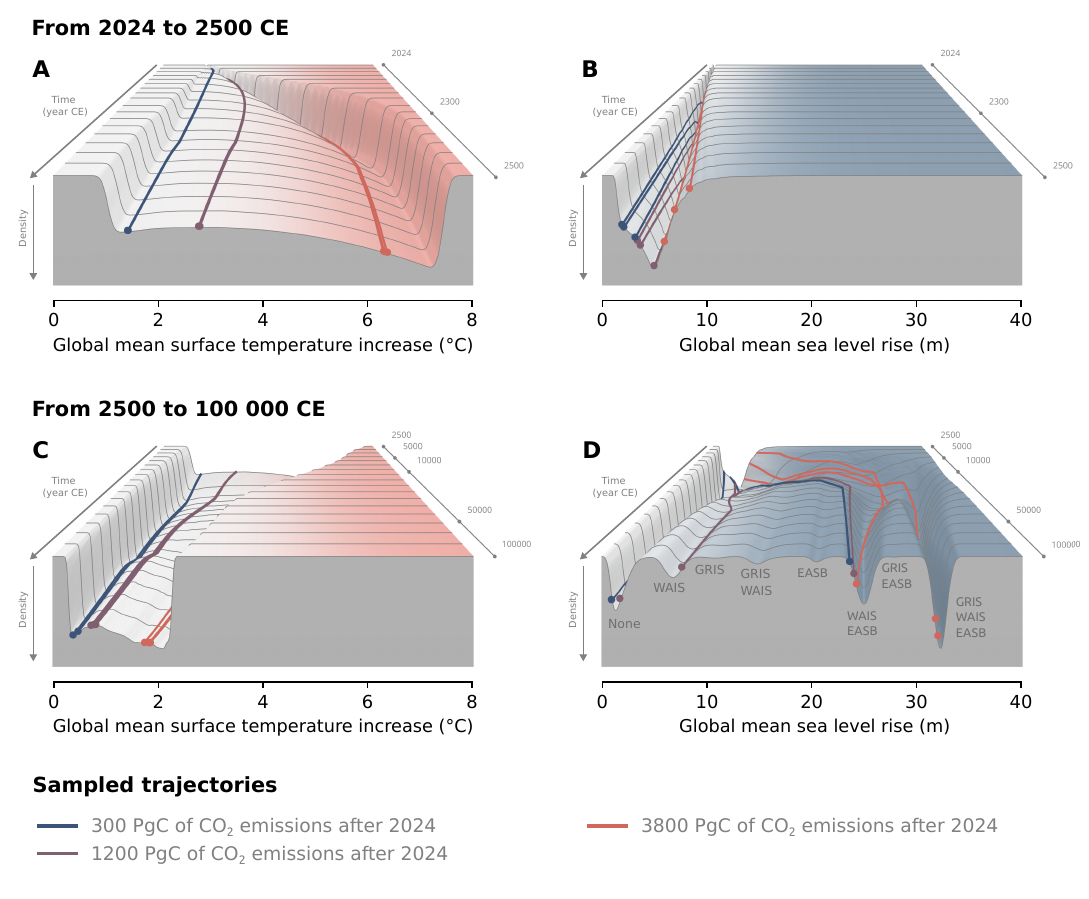}
    \caption{\textbf{Evolution of GMST and GMSL distributions across scenarios and model runs.} \textbf{(A, C)} Global mean surface temperature (GMST) and \textbf{(B, D)} global mean sea level (GMSL) distributions as functions of time in the coupled experiment. The time axis is logarithmic, and distributions are normalized independently to their maximum value at each time $t$. Colored lines show trajectories for three parameter sets under a low-emission scenario (300~PgC), a medium-emission scenario (1200~PgC), and a high-emission scenario (3800~PgC). In panel (D), acronyms indicate which elements tipped among the Greenland Ice Sheet (GRIS), the West Antarctic Ice Sheet (WAIS), and the East Antarctic Subglacial Basins (EASB). This figure is adapted from \cite{steffen_trajectories_2018}; a detailed comparison is provided in the supplementary text.}
    \label{fig:distributions_3D}
\end{figure}


\clearpage 

%
\bibliography{submission_Science/science_template} 
\bibliographystyle{sciencemag}

%
%
%
%
%
%


\section*{Acknowledgments}
We acknowledge the Global Carbon Project, which is responsible for the Global Carbon Budget and we thank the different modelling groups for producing and making available their model output. The scientific colormap lipari \cite{crameri_scientific_2023} is used in most figures. The authors used a language model (ChatGPT, OpenAI) to assist with minor text editing and phrasing improvements.

\medskip

\noindent\textit{We dedicate this work to the memory of Dr. Will Steffen, an outstanding scientist whose commitment to understanding and preserving the Earth system remain a lasting inspiration for climate science and this study in particular.}

\paragraph*{Funding:}
Most of this work was carried out while V.~C was supported by the Belgian National Fund for Scientific Research (F.R.S.–FNRS) through an Aspirant Fellowship (FC~38941). V.~C also received support from UCLouvain. W.~T and V.~C. acknowledge funding from the European Research Council (ERC) under the European Union's Horizon Framework research and innovation programme (grant agreement No~101076909; ERC Consolidator Grant ‘LACRIMA’). 

\paragraph*{Author contributions:}
V.~C and M.~C conceptualized the project and developed the methodology. M.~C supervised the project. V.~C implemented the model, performed the simulations, and created the visualizations. V.~C wrote the original draft. W.~T and M.~C reviewed and edited the manuscript, providing detailed feedback.

\paragraph*{Competing interests:}
There are no competing interests to declare.
\paragraph*{Data and materials availability:}
 Historical CO$_2$
 emissions are from \cite{friedlingstein_global_2025} and are available at \url{https://globalcarbonbudgetdata.org/latest-data.html} (accessed 10 September 2025). Historical CH$_4$ emissions are from \cite{jones_national_2023} and are available at \url{https://zenodo.org/records/10839859} (accessed 10 September 2025). CO$_2$ and CH$_4$ emissions for the SSP scenarios are available from \url{https://tntcat.iiasa.ac.at/SspDb} (accessed 10 September 2025). Sea-level rise data used in Fig. S8 originate from IPCC AR6 WG1 Chapter 9 \cite{fox-kemper_ocean_2021} and are publicly available at \url{https://github.com/BrodiePearson/IPCC_AR6_Chapter9_Figures/tree/main/Plotting_code_and_data/Fig9_27_SL_scenarios} (accessed 12 December 2025). Sea-level rise data used in Fig. S9 are from \cite{clark_consequences_2016}, available at \url{https://www.nature.com/articles/nclimate2923#Sec6}
 (accessed 12 December 2025); from \cite{van_breedam_semi-equilibrated_2020}, available from the authors upon request; and from \cite{garbe_hysteresis_2020}, available from the authors upon request. The version of SURFER used in this study, together with the input files required to run the model and the data used to generate most figures, is archived on Zenodo \cite{couplet_surferv30TEs_2025,couplet_2025_inputdata}. SURFER is released under the MIT license.


\subsection*{Supplementary materials}
Supplementary Text\\
Figs. S1 to S10\\
Tables S1 to S6\\
References \textit{(42-\arabic{enumiv})}\\ 


\newpage


\renewcommand{\thefigure}{S\arabic{figure}}
\renewcommand{\thetable}{S\arabic{table}}
\renewcommand{\theequation}{S\arabic{equation}}
\renewcommand{\thepage}{S\arabic{page}}
\setcounter{figure}{0}
\setcounter{table}{0}
\setcounter{equation}{0}
\setcounter{page}{1} 


\begin{center}
\section*{Supplementary Materials for\\ \scititle}

Victor~Couplet$^{*}$,
Wim~Thiery,
Michel~Crucifix\\ 
\small$^\ast$Corresponding author. Email: victor.couplet@vub.be\\
\end{center}

\subsubsection*{This PDF file includes:}
Supplementary Text\\
Figures S1 to S10\\
Tables S1 to S6\\

\newpage



\subsection*{Supplementary Text}

\subsubsection*{Stability landscapes versus density of trajectories}
\label{appendix:landscapes}

In \cite{steffen_trajectories_2018}, the hothouse hypothesis in summarized in a conceptual figure (reproduced here, Fig. \ref{fig:stability_vs_density}A) which illustrates two divergent pathways within an evolving stability landscape : a “Hothouse Earth” trajectory driven by reinforcing biogeophysical feedbacks once a planetary threshold is crossed, and a “Stabilized Earth” pathway that avoids this threshold through sustained and deliberate human interventions. Because our work aims to test the Hothouse Earth hypothesis, we intentionally designed Figure 5 to resemble the figure from \cite{steffen_trajectories_2018} for comparability. However, these figures present some important differences, which we detail below. 

First, the figure proposed by \cite{steffen_trajectories_2018} is an illustrative “artist’s view”, whereas our figure is derived from model data, allowing for a fully quantitative representation. Along the x-axis, we plot global mean temperature (Fig. \ref{fig:distributions_3D}, A and C) or global mean sea level (GMSL; Fig. \ref{fig:distributions_3D}, B and D) rather than an abstract variable representing the overall state of the Earth System. The y-axis in our figure explicitly represents time, providing temporal context. In the z-dimension, \cite{steffen_trajectories_2018} depicts stability as the inverse of potential energy. However, potential energy is only well-defined in conservative systems, which the Earth System is not. For this reason, we represent stability using the density of trajectories across all our model simulations. High-density regions indicate states that frequently occur in SURFER across multiple emissions scenarios and parameter sets, and can therefore be interpreted as likely stable. 

The qualifier “likely” is essential: while stable states will generate high-density regions because they act as attractors, the reverse is not always true. Density can also reflect the distribution of initial conditions and prescribed emissions. For example, in Figure \ref{fig:distributions_3D}A, between 2024 and 2300, densities are higher at warmer temperatures, and one might misinterpret this as evidence that warmer states are inherently more stable over short timescales. In reality, this pattern reflects the skewed distribution of transient cumulative emissions in our prescribed scenario set, which is biased toward higher emissions during this period (Fig. \ref{fig:distributions_scenarios}). A second factor influencing density patterns is the choice of initial conditions. Because all our simulations start from the same initial state, the initial distribution is concentrated around that point, masking alternative stable states that would appear in a stability-landscape representation. For instance, the sea-level rise modes associated with the collapsed state of the ice-sheets are absent early on, despite being possible equilibrium states under low temperatures (ice sheets are represented as bi-stable systems in our models, so both the collapsed and non-collapsed state coexist under a range of temperatures). Similarly, if SURFER were able to simulate glacial–interglacial limit cycles, they would not appear in our figure for the shorter time scales. 

Another fundamental difference lies in how human societies are included. The framing in \cite{steffen_trajectories_2018} considers human societies as an integral and interacting component of the Earth system. This holistic view is valuable for highlighting governance challenges and rethinking humanity’s role in the Earth system, but it comes at the cost of analytical clarity. Indeed, blending together biogeophysical and societal processes within a single metaphorical stability landscape leads to some ambiguities. For instance, if the hypothesized planetary threshold is a property of the physical Earth System, defined by cascading tipping elements, it should separate two intrinsic states: a “Hothouse Earth” and a “Stabilized Earth.” Yet \cite{steffen_trajectories_2018} argues that the latter likely depends on continuous human feedbacks (e.g., emissions reductions, carbon sink enhancement, geoengineering). If interpreted this way, and permanent intervention is required to prevent a transition to a Hothouse Earth, then the Earth system has effectively already crossed a physical threshold, contradicting their definition of a “Stabilized Earth” pathway as one that avoids such a threshold. These technical nuances, however, do not weaken their central point: collective human action is indispensable to maintaining a habitable Earth system.

In our framework, by contrast, we treat the Earth System and human societies separately, with the latter represented as an external forcing through prescribed emissions scenarios. Our figure represents a range of possible Earth system trajectories, independent of assumptions about the most probable emission pathway or interventions that depend on the system’s evolving state. Nevertheless, governance options can still be explored by comparing distributions for a subset of emission scenarios, thereby assessing how mitigating CO2 emissions can reduce the risk of following undesirable trajectories (Fig. \ref{fig:distributions_2D} and Table \ref{tab:scenario_modes}). 

A final difference lies in the types of human interventions considered. In our framework, human societies can only act by controlling their CO$_2$ emissions. When we say that tipping cascades may \textit{lock} the Earth system on a pathway towards a melthouse, we mean that it is the long-term committed state based on a given emission scenario, and without any further human intervention. In contrast, \cite{steffen_trajectories_2018} considers a range of other interventions such as the enhancement or restoration of existing carbon sinks, carbon dioxide removal, and solar radiation modification. Whether these Earth system management strategies could avert a hothouse or melthouse outcome under specific emissions scenarios could be systematically investigated using SURFER, building on the methodology developed in \cite{martinez_montero_lost_2024}.

\subsubsection*{Sea-level rise in SURFER}
\label{appendix:slr}

Global mean sea level rise is estimated as the sum of 6 contributions: thermal expansion ($S_{\text{th}}$), and melt from the mountain glaciers ($S_{\text{gl}}$), the Greenland Ice Sheet ($S_{\text{GRIS}}$), the West and East Antarctic Ice Sheets ($S_{\text{WAIS}}$ and $S_{\text{EAIS}}$ ), and the East Antartic Sublgacial Basins ($S_{\text{EASB}}$)  :
\begin{equation}
S_{\mathrm{tot}}=S_{\mathrm{th}}+S_{\mathrm{gl}}+S_{\mathrm{GRIS}}+S_{\mathrm{WAIS}}+S_{\mathrm{EASB}}+S_{\mathrm{EAIS}}\,.
\end{equation}
Thermal expansion is given by 
\begin{equation}
S_{\mathrm{th}}= \alpha_{\mathrm{U}} h_{\mathrm{U}} \delta T_{\mathrm{U}} + \alpha_{\mathrm{I}} h_{\mathrm{I}} \delta T_{\mathrm{I}} + \alpha_{\mathrm{D}} h_{\mathrm{D}} \delta T_{\mathrm{D}}\,,
\end{equation}
where $\alpha_i$ is the thermal expansion coefficient corresponding to ocean layer $i\in\{\mathrm{U,I,D}\}$.

The evolution of the sea-level rise contribution from glaciers is the same as in previous SURFER versions and is given by the equation
\begin{equation}
\frac{\mathrm{d} S_{\mathrm{gl}}}{\mathrm{d} t}=\frac{1}{\tau_{\mathrm{gl}}}\left(S_{\mathrm{gleq}}\left(\delta T_{\mathrm{U}}\right)-S_{\mathrm{gl}}\right)\,,
\end{equation}
with 
\begin{equation}
S_{\mathrm{gleq}}\left(\delta T_{\mathrm{U}}\right)=S_{\mathrm{gl} \text { pot }} \tanh \left(\frac{\delta T_{\mathrm{U}}}{\zeta}\right)\,.
\end{equation}
Here $\tau_{\mathrm{gl}}$ is a relaxation timescale, $S_{\mathrm{gl} \text { pot }}$ is the potential sea level rise due to mountain glaciers and $\zeta$ is a sensitivity coefficient. 

The contributions from the ice sheets are given by
\begin{equation}
    S_{i} = S^{\text{pot}}_i\left(1-x_{\mathrm{i}}(t)\right), \quad i \in \{\text{GRIS, WAIS, EASB, EAIS}\},
\end{equation}
where $S_{i}$ is the sea-level rise potential of the ice sheet. The value of the parameters from the equations above are given in Table \ref{tab:slrparams_1}.

Projections of sea-level rise from earlier SURFER versions were successfully calibrated against higher complexity models. SURFER v2.0, which excludes long-term carbon cycle processes such as sedimentation and weathering, simulated short-term sea-level rise within the range of IPCC projections \cite{martinez_montero_surfer_2022}. In \cite{couplet_surfer_2025}, we demonstrated that SURFER v3.0 produces multi-millennial thermosteric sea-level rise comparable to other EMICs and showed that incorporating long-term carbon cycle processes reduces the contributions from the Greenland and Antarctic ice sheets. Here, we demonstrate that splitting the Antarctic Ice Sheet into the WAIS, EASB, and EAIS components still results in reasonable sea level rise projections. 

Figure \ref{fig:slr_ipcc} compares SURFER's simulated sea-level rise under SSP1-2.6 and SSP5-8.5 emission scenarios with IPCC projections. For this comparison, we exclude interactions between tipping elements and their feedbacks on temperature. However, we account for the uncertainty associated with tipping elements by running the model across our ensemble of parameter sets (decoupled experiment settings). SURFER's projections align well with the IPCC's for both scenarios: the median values are comparable, and SURFER's likely range generally falls within the IPCC's medium-confidence likely range. Beyond 2090, even under worst-case scenarios, SURFER does not simulate sea-level rise as high as the IPCC's low-confidence 95th percentile. Note that the IPCC's uncertainty range reflects a combination of uncertainties in both equilibrium climate sensitivity (ECS) and ice sheet dynamics. In contrast, for this figure, SURFER's ECS is fixed at 3.5°C, and its uncertainty range reflects only uncertain ice sheet dynamics. This explains why SURFER's range is narrower than that of the IPCC.

In Figure \ref{fig:slr_commitment}, we compare SURFER's 10,000-year sea-level rise commitment with projections from \cite{clark_consequences_2016} and \cite{van_breedam_semi-equilibrated_2020}. We examine sea-level rise as a function of both peak warming, following the approach in the sixth IPCC report [see Fig. 9.30 \cite{ipcc_climate_2021}], and long-term warming. SURFER simulates lower sea-level rise contributions from ice sheets for the same peak warming than the other models, resulting in an overall lower sea-level rise commitment. This difference arises because, in SURFER, temperatures decrease more rapidly after peaking than in UVic and LOVECLIM, the EMICs used in the studies referenced. However, when we evaluate sea-level commitment as a function of long-term warming, SURFER aligns with the other models, suggesting that the ice sheets are reasonably well-calibrated. This also highlights that a significant reduction in CO$_2$ concentrations over millennia, whether natural or through carbon dioxide removal, could help reduce committed sea-level rise, as demonstrated in \cite{couplet_surfer_2025}.

\subsubsection*{Sensitivity analysis}
\label{appendix:sensitivity}

In the main text, we presented two experiments: the decoupled experiment, which excludes interactions between elements and their feedback on temperature, and the coupled experiment, which incorporates both. We conducted both experiments using an equilibrium climate sensitivity of 3.5°C. This section examines how our results change under a higher ECS and stronger interaction strengths, focusing specifically on the risk of reaching a hothouse or melthouse state, as illustrated in Figure \ref{fig:hothouse_risk_sensitivity}.

The risk of reaching a hothouse state is higher for a high-end ECS of 5°C compared to an ECS of 3.5°C and becomes non-negligible ($>\sim$10\%) for cumulative emissions around 1450~PgC. With an ECS of 5°C, the risk of entering a melthouse state becomes significant even under the 300~PgC low-emission scenario, rising to 15\%, compared to just 1\% for an ECS of 3.5°C.

Doubling all interaction strengths between tipping and other non-linear elements while maintaining an ECS of 3.5°C results in a negligible increase in the risk of reaching a hothouse state compared to the standard coupled experiment (purple and black lines in Figure \ref{fig:hothouse_risk_sensitivity} overlap). Although more elements tip due to stronger interactions, their influence on maximum temperatures remains minimal. However, these elements can significantly affect sea level rise, increasing the risk of entering a melthouse state.

These findings, along with those presented in the main text, suggest that emission pathways and the climate response primarily dictate short-term and peak warming and, consequently, the risk of reaching a hothouse state. In comparison, the influence of non-linear elements and cascading dynamics appears relatively minor. Long-term temperatures are additionally shaped by millennial-scale carbon cycle processes such as sedimentation and weathering.

Short-term sea level rise is primarily driven by near-term temperatures and the characteristics of individual ice sheets. In contrast, long-term and peak sea level rise, which determine the risk of reaching a melthouse state, are determined by the long-term temperatures and the properties of ice sheets, including their interactions with other elements of the Earth system.

It is important to note that the temperature response for a given CO$_2$ emissions scenario—and thus the risk of reaching a hothouse or melthouse state — also depends on other greenhouse gases, such as methane, and forcing agents like aerosols. In this study, methane emissions are set to the levels of the extended SSP1-2.6 scenario for all CO$_2$ emissions pathways \cite{meinshausen_shared_2020}, while aerosol forcing is assumed to be zero throughout.

\newpage


\begin{figure}
\centering
\includegraphics[width=0.8\textwidth]{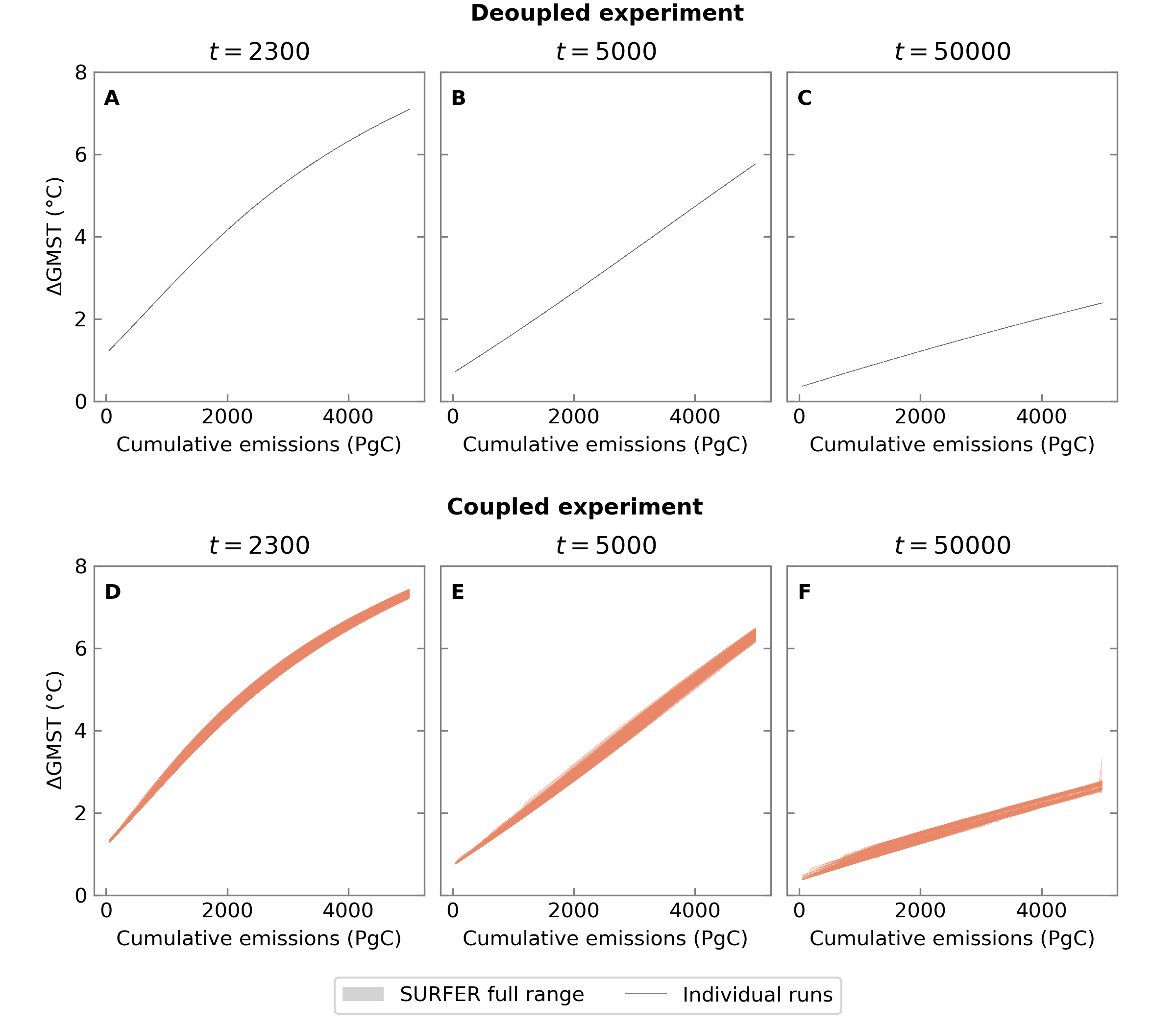}
\caption{\textbf{Global mean surface temperature (GMST) at different times as a function of cumulative emissions, relative to the year 1850 CE.} Results are shown for the decoupled \textbf{(A-C)} and coupled \textbf{(D-E)} experiments. Shaded areas show the full range across all simulations and thinner solid lines represent individual runs. In the decoupled experiment, GMST is independent of the states of tipping and threshold-free elements, producing indistinguishable results across the 1,000 parameter sets in panels (A-C). }
\label{fig:impacts_times_T}
\end{figure}

\begin{figure}
\centering
\includegraphics[width=0.8\textwidth]{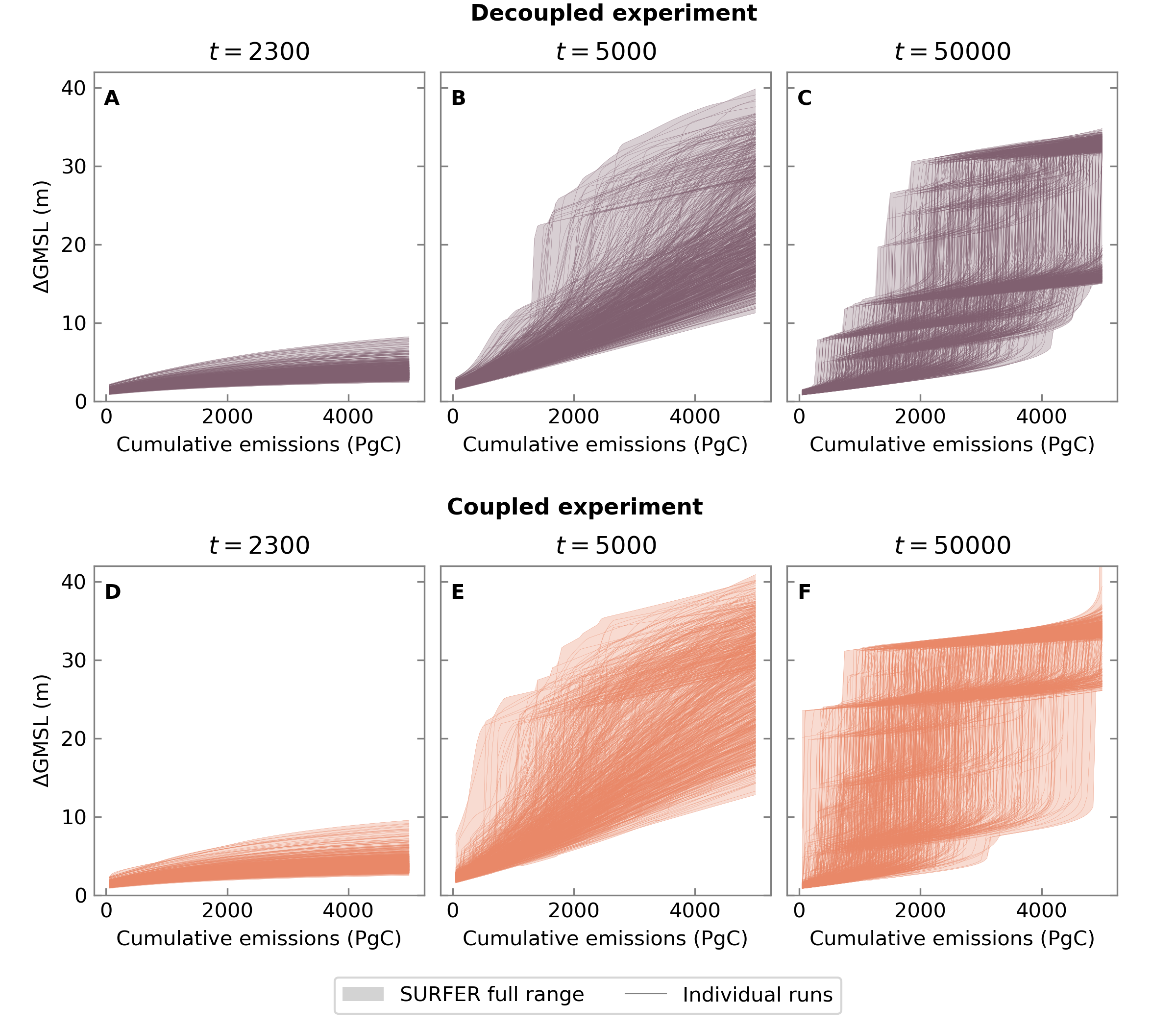}
\caption{\textbf{Global mean sea level (GMSL) at different times as a function of cumulative emissions, relative to the year 1850 CE.} Results are shown for the decoupled \textbf{(A-C)} and coupled \textbf{(D-E)} experiments. Shaded areas show the full range across all simulations and thinner solid lines represent individual runs.}
\label{fig:impacts_times_slr}
\end{figure}

\begin{figure}
\centering
\includegraphics[width=0.8\textwidth]{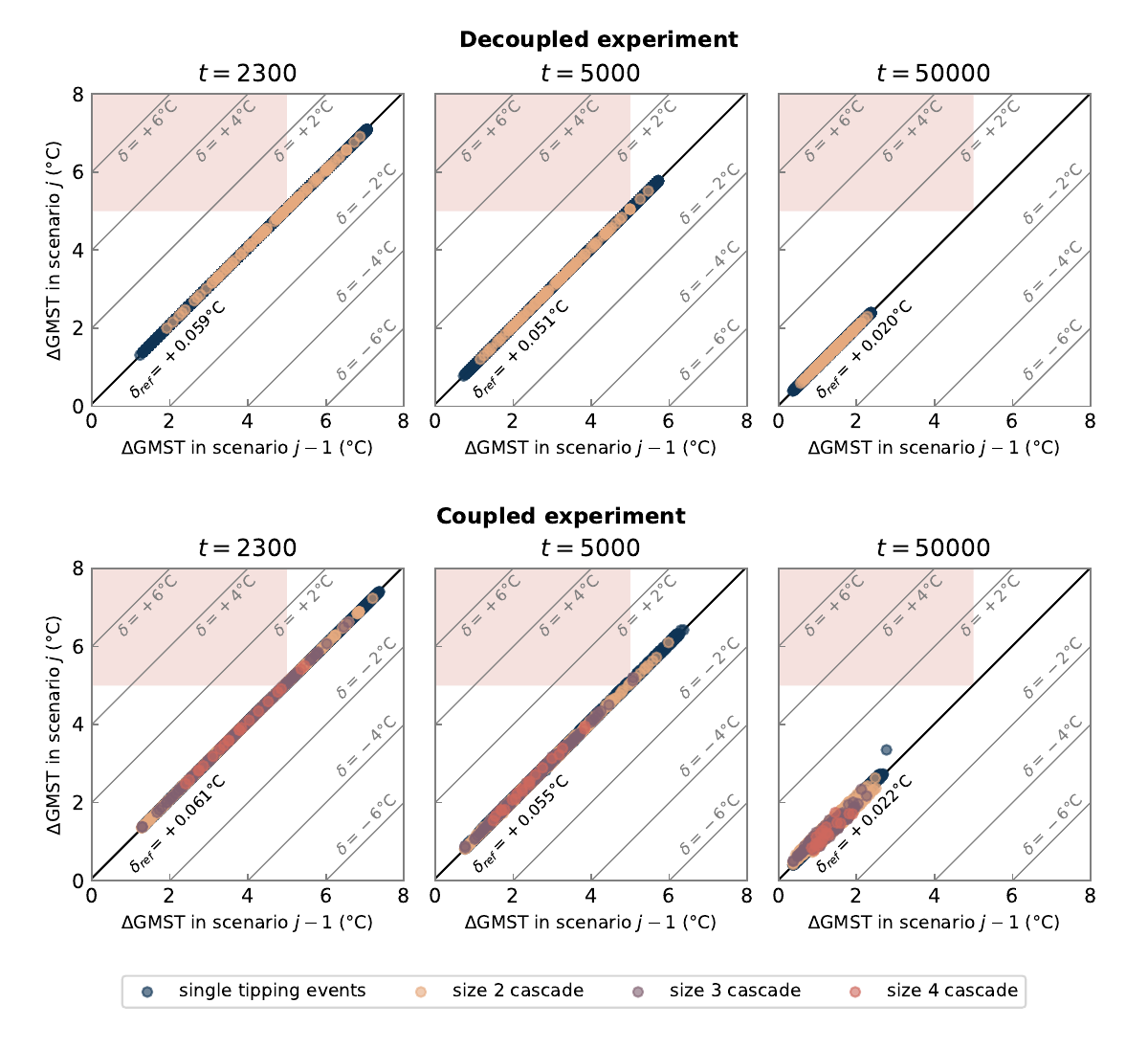}
\caption{\textbf{Influence of tipping events on global mean temperature in 2300 CE, 5000 CE, and 50000 CE for the decoupled and coupled experiments.} Each observed tipping event is shown as a point, colored by event type. The temperature impact of each event is assessed by comparing the coordinates of the corresponding point, which represent the warming in scenario $j$ (where the event occurred) and in scenario $j-1$ (50~PgC lower cumulative emissions). The thick black diagonal in each plot has equation $y=x+\delta_{ref}$, where $\delta_{ref}$ is the mean difference in warming between adjacent scenarios without tipping events. Points on this line have a negligible influence, whereas deviations from it indicate the strength of the tipping event’s effect. The thin gray diagonals indicate contours of equal warming response and are provided as visual guides. Shaded regions highlight tipping events that push warming past 5°C above pre-industrial levels (1850~CE), starting from below this threshold. In the decoupled experiment, the interactions between elements are disabled, so events identified as size-2 cascades arise only from coincident tipping of independent elements.}
\label{fig:gaps_times_T}
\end{figure}

\begin{figure}
\centering
\includegraphics[width=0.8\textwidth]{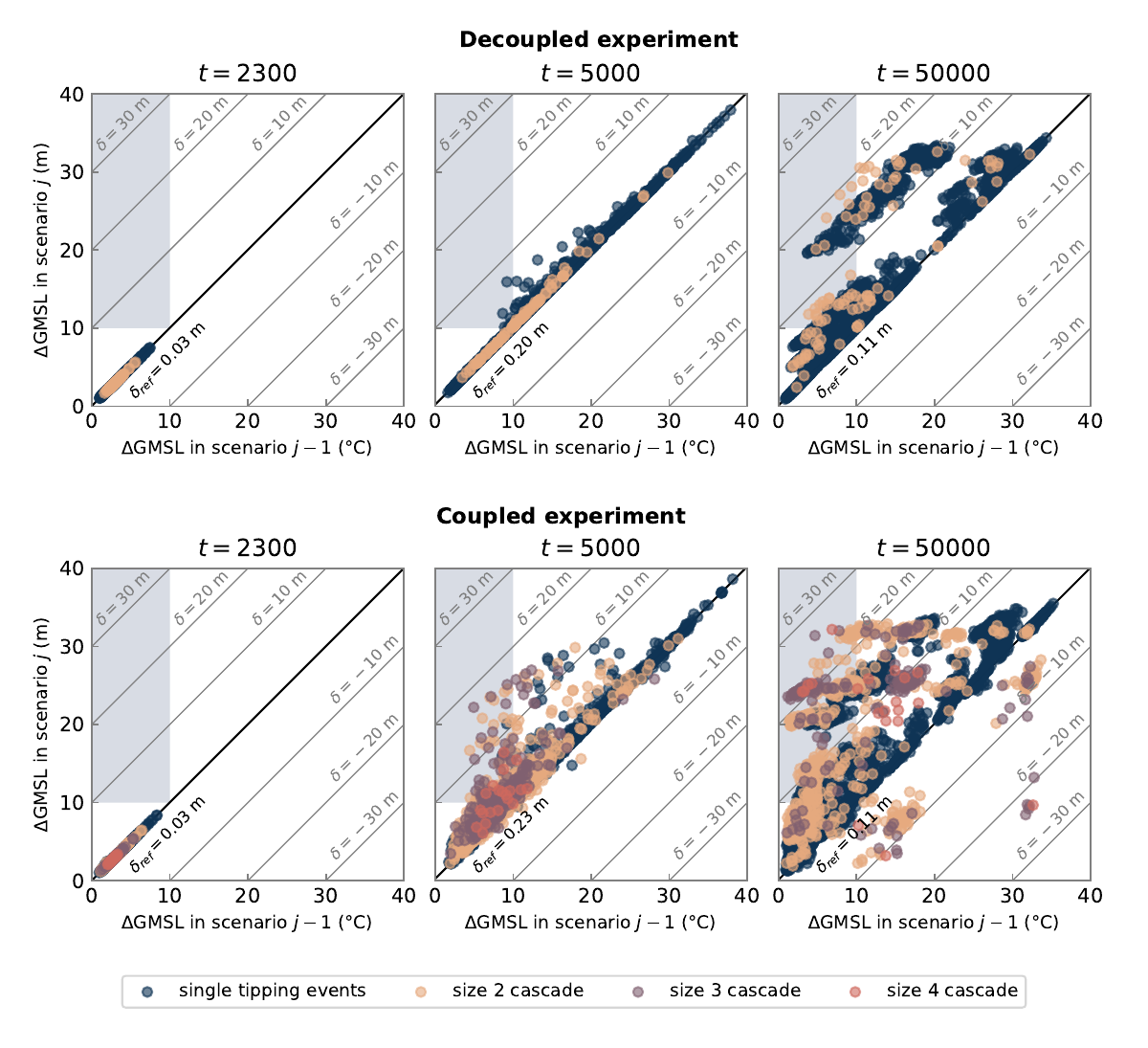}
\caption{\textbf{Influence of tipping events on global mean sea level (GMSL) in 2300 CE, 5000 CE, and 50000 CE for the decoupled and coupled experiments.} Each observed tipping event is shown as a point, colored by event type. The GMSL impact of each event is assessed by comparing the coordinates of the corresponding point, which represent the peak sea level rise in scenario $j$ (where the event occurred) and in scenario $j-1$ (50~PgC lower cumulative emissions). The thick black diagonal in each plot has equation $y=x+\delta_{ref}$, where $\delta_{ref}$ is the mean difference in sea level rise between adjacent scenarios without tipping events. Points on this line have a negligible influence, whereas deviations from it indicate the strength of the tipping event’s effect. The thin gray diagonals indicate contours of equal sea level rise response and are provided as visual guides. Shaded regions highlight tipping events that cause sea level to rise by more than 10~m above pre-industrial levels (1850~CE), starting from below this threshold. In the decoupled experiment, the interactions between elements are disabled, so events identified as size-2 cascades arise only from coincident tipping of independent elements.}
\label{fig:gaps_times_slr}
\end{figure}

\begin{figure}
    \centering
    \includegraphics[width=\textwidth]{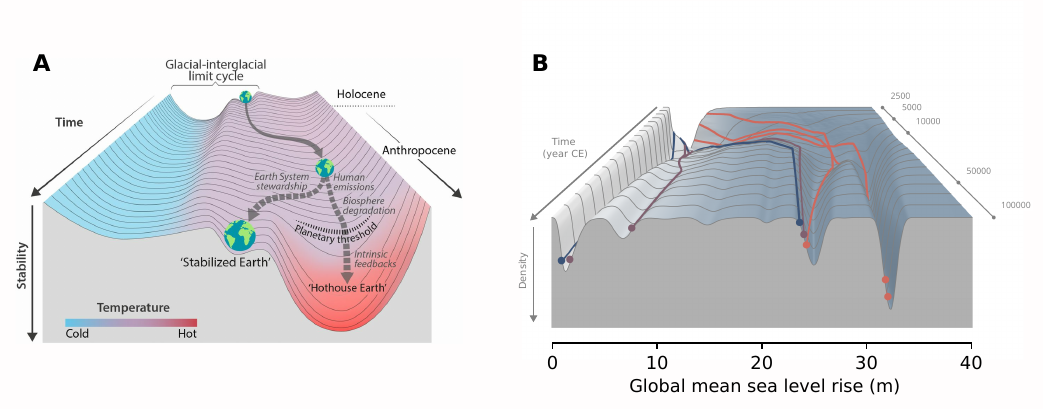}
    \caption{\textbf{Comparison between a conceptual stability landscape and our representation of Earth system stability through time evolving distributions.} \textbf{(A)} Conceptual stability landscape reproduced from Steffen et al. (2018), PNAS. © 2018 The Author(s), CC BY-NC-ND 4.0. \textbf{(B)} Time evolving distribution of sea level rise across all scenarios and model runs. This is a reproduction of Figure \ref{fig:distributions_3D}D in the main text.}
    \label{fig:stability_vs_density}
\end{figure}

\begin{figure}
\centering
\includegraphics[width=0.8\textwidth]{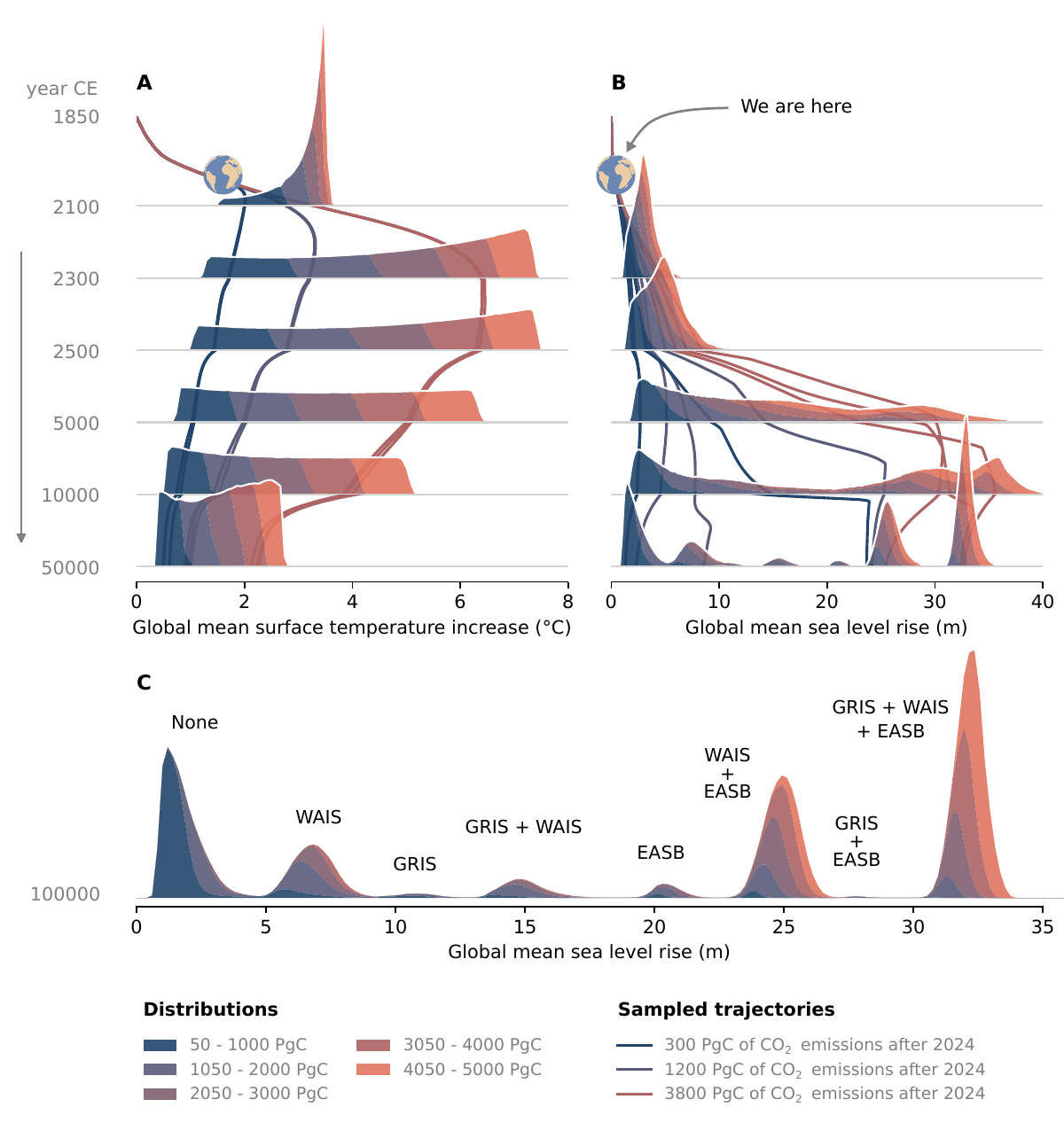}
\caption{\textbf{Evolution of GMST and GMSL distributions across scenarios and model runs.} \textbf{(A)}~Global mean surface temperature (GMST) and \textbf{(B,C)} global mean sea level (GMSL) distributions as functions of time in the coupled experiment. The time axis is on a linear scale with several scale breaks. The temperature distribution at $t=2100$ and the sea level rise distributions at $t=2300$ and $t=2300$ have been scaled down to fit the figure. The sea level rise distribution at $t=2100$ is too narrow and is not plotted. Colored lines illustrate trajectories for three parameter sets under a low-emission scenario (300~PgC), a medium-emission scenario (1200~PgC), and a high-emission scenario (3800~PgC). In panel (C), acronyms indicate which elements tipped among the Greenland Ice Sheet (GRIS), the West Antarctic Ice Sheet (WAIS) and the East Antarctic Subglacial Basins (EASB).}
\label{fig:distributions_2D}
\end{figure}

\begin{figure}
\centering
\includegraphics[width=0.6\textwidth]{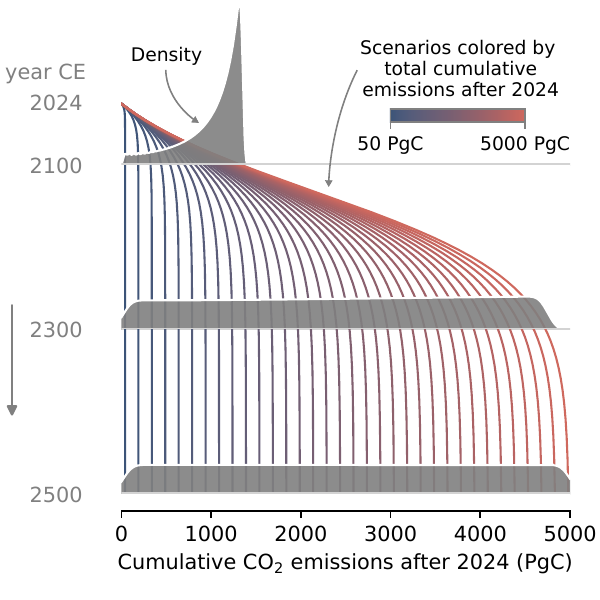}
\caption{\textbf{Distribution of transient cumulative emissions across the 100 scenarios.} Although the total post-2024 cumulative emissions in each scenario are uniformly distributed between 50 and 5000~PgC, the transient cumulative emissions exhibit a non-uniform distribution.}\label{fig:distributions_scenarios}
\end{figure}
\clearpage

\begin{figure}
\centering
\includegraphics[width=0.8\textwidth]{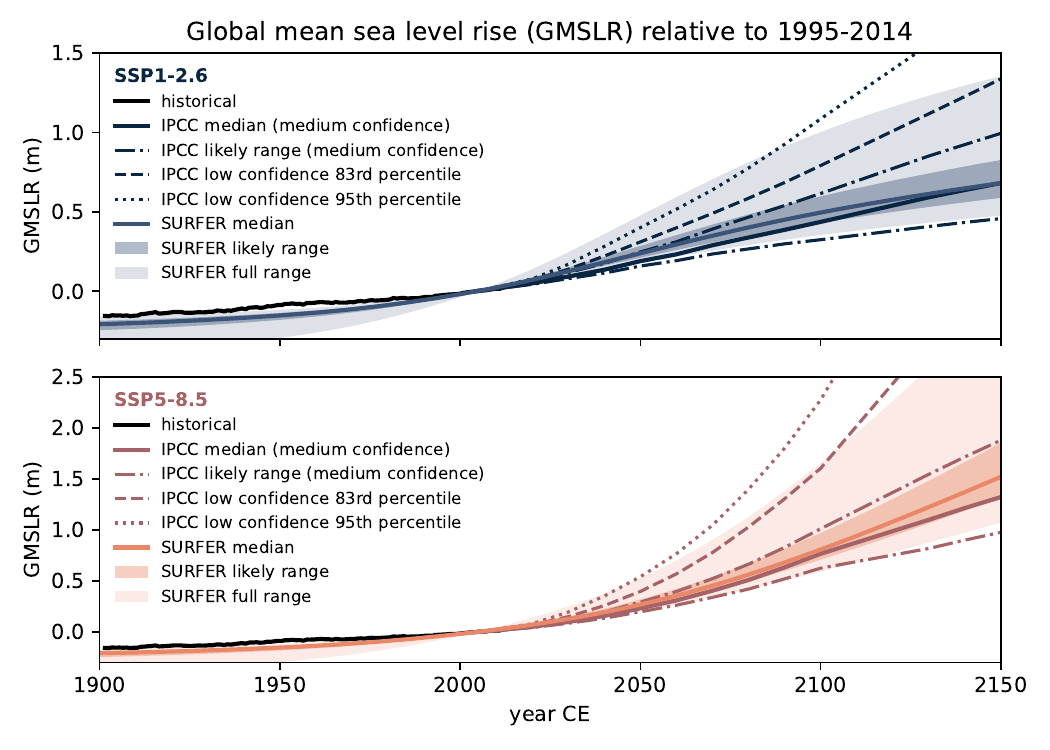}
\caption{\textbf{Sea level rise projections from SURFER compared with IPCC estimates.} Projections are shown for the SSP1-2.6 and SSP5-8.5 scenarios. Likely ranges correspond to the 17th–83rd percentiles and are indicated by dash-dotted lines for the IPCC and shaded regions for SURFER. SURFER’s results are shown for the decoupled experiment, where uncertainty arises solely from parameter choices in calibrating the ice sheets, with equilibrium climate sensitivity fixed at 3.5°C. This figure is adapted from figure 9.27 of the IPCC AR6 WG1 report \cite{ipcc_climate_2021}.}
\label{fig:slr_ipcc}
\end{figure}

\begin{figure}
\centering
\includegraphics[width=\textwidth]{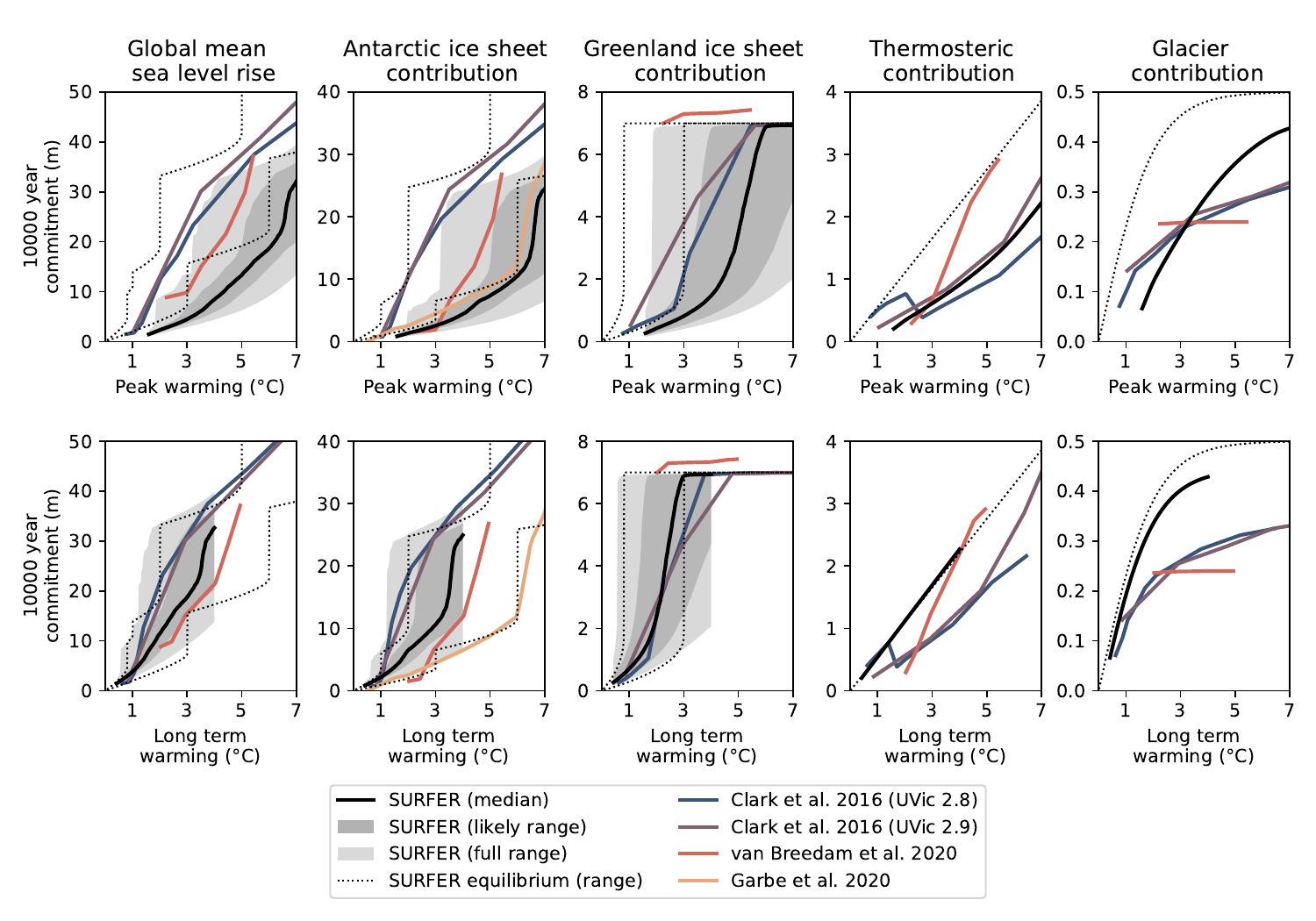}
\caption{\textbf{Long-term sea-level rise commitment as a function of peak and long-term warming}. The commitment is calculated as the difference in sea level between the years 12,000 CE and 2000 CE. Temperature changes are expressed relative to the 1850–1900 CE baseline. In the bottom panels, long-term warming refers to the GMST anomaly in 12,000 CE. The leftmost column shows the total global mean sea-level rise, while the other columns show contributions from individual sources. Coloured lines represent projections from various models \cite{clark_consequences_2016, van_breedam_semi-equilibrated_2020, garbe_hysteresis_2020}, and the black line represents SURFER's median outcome for the decoupled experiment. Shaded regions show SURFER's likely range (13th to 83rd percentile) and full range. The Antarctic Ice Sheet contribution in \cite{garbe_hysteresis_2020} is computed from quasi-equilibrium runs. This figure is adapted from Figure 9.30 of the IPCC AR6 WG1 report \cite{ipcc_climate_2021}.\label{fig:slr_commitment}}
\end{figure}

\begin{figure}
\centering
\includegraphics[width=0.6\textwidth]{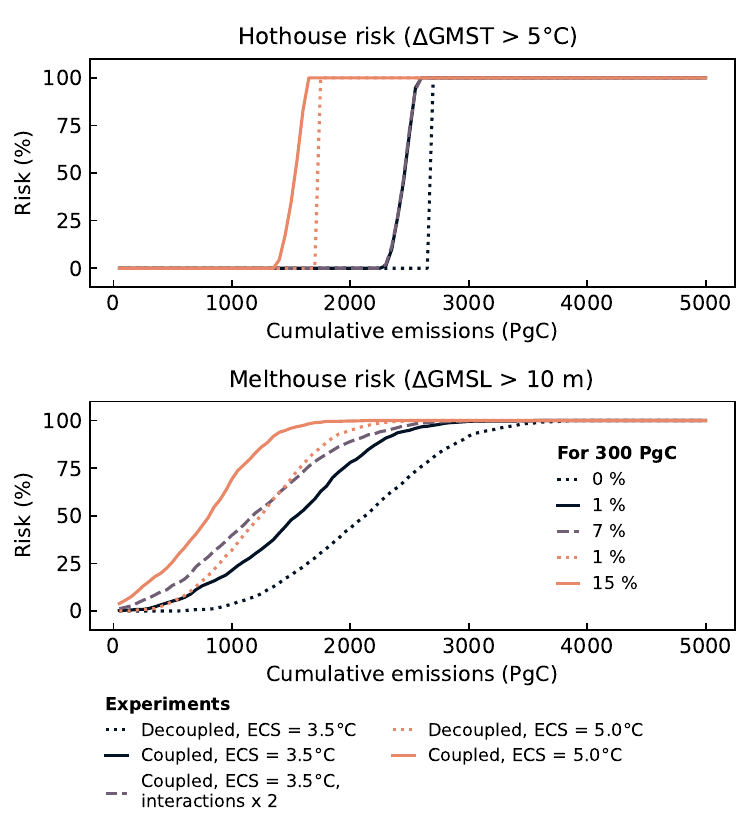}
\caption{\textbf{Sensitivity of hothouse and melthouse risks to model configuration.} Risk is shown as a function of cumulative emissions, defined as the total emitted after 2024 CE. For each experiment, the risk corresponds to the percentage of ensemble simulations, with varying parameter sets, that exceed the reference thresholds ($\Delta \text{GMST} > 5~^\circ C$ and $\Delta \text{GMSL} > 10~m$) at any given time.}
\label{fig:hothouse_risk_sensitivity}
\end{figure}


\begin{table}
    \centering
    \caption{\textbf{Sampled critical temperatures and internal time scales of tipping and other non-linear threshold-free elements for the example in Figure 1.}}
    \label{tab:elements_params_sampled}
    \begin{tabular}{lccr}
    \\
    \hline
    Tipping      & $T_+$   & $\tau_-$   & $\tau_+$ \\
    elements     & (°C)    & (years)    & (years)  \\
    \hline
    GRIS         & $1.95$   & $1121.7$   & $5500.0$ \\
    WAIS         & $1.15$   & $241.3$   & $5500.0$ \\
    EASB         & $3.86$   & $313.5$    & $5500.0$ \\
    EAIS         & $9.26$   & $560.1$    & $5500.0$ \\
    AMOC         & $1.42$   & $19.5$     & $19.5$   \\
    AMAZ         & $5.93$   & $10.1$     & $10.1$   \\
    \hline
    Other        & $T_I$   & $\tau_-$   & $\tau_+$ \\
    elements     & (°C)    & (years)    & (years)  \\
    \hline
    ASI          & $5.43$   & $6.8$      & $6.8$    \\
    PERM         & $4.02$   & $36.1$     & $3000.0$ \\
    \hline
    \end{tabular}
\end{table}

\begin{table}
    \centering
    \caption{\textbf{Sampled interactions parameters for the example in Figure 1.} The parameter $\epsilon_{j,i}$ describes the influence of element $i$ on element $j$.}
    \label{tab:interaction_params_sampled}
    
    \begin{tabular}{llcccccccc}
    \\
    \hline
    Elements &   & $\epsilon_{\text{GRIS},i}$ & $\epsilon_{\text{WAIS},i}$ & $\epsilon_{\text{EASB},i}$ & $\epsilon_{\text{EAIS},i}$ & $\epsilon_{\text{AMOC},i}$ & $\epsilon_{\text{AMAZ},i}$ & $\epsilon_{\text{ASI},i}$ & $\epsilon_{\text{PERM},i}$ \\
    \hline
    GRIS  &   & $0.00$  & $0.41$  & $0.36$  & $0.00$  & $3.79$  & $0.00$  & $0.00$  & $0.00$ \\
    WAIS  &   & $0.26$  & $0.00$  & $0.00$  & $0.00$  & $5.72$  & $0.00$  & $0.00$  & $0.00$ \\
    EASB  &   & $0.24$  & $0.00$  & $0.00$  & $0.00$  & $0.00$  & $0.00$  & $0.00$  & $0.00$ \\
    EAIS  &   & $0.00$  & $0.00$  & $0.00$  & $0.00$  & $0.00$  & $0.00$  & $0.00$  & $0.00$ \\
    AMOC  &   & $-0.98$ & $0.66$  & $0.94$  & $0.00$  & $0.00$  & $-0.21$  & $-0.64$ & $0.00$ \\
    AMAZ  &   & $0.00$  & $0.00$  & $0.00$  & $0.00$  & $0.00$  & $0.00$  & $0.00$  & $0.00$ \\
    ASI   &   & $0.28$  & $0.00$  & $0.00$  & $0.00$  & $0.55$  & $0.00$  & $0.00$  & $0.16$ \\
    PERM  &   & $0.00$  & $0.00$  & $0.00$  & $0.00$  & $0.31$  & $0.00$  & $0.00$  & $0.00$ \\
    \hline
    \end{tabular}
\end{table}

\begin{table}
    \centering
    \caption{\textbf{Sampling ranges for the critical temperature and internal time scales of tipping and other non-linear threshold-free elements.}}
    \label{tab:elements_params_reduced}
    
    \begin{tabular}{lccr}
    \\
    \hline
    Tipping      & $T_+$         & $\tau_-$     & $\tau_+$ \\
    elements     & (°C)          & (years)      & (years)  \\
    \hline
    GRIS         & $[0.8,3.0]$   & $[306,1981]$ & $5500$    \\
    WAIS         & $[1.0,3.0]$   & $[153,1717]$ & $5500$    \\
    EASB         & $[2.0,6.0]$   & $[88,839]$   & $5500$    \\
    EAIS         & $[5.0,10.0]$  & $[470,741]$  & $5500$    \\
    AMOC         & $[1.4,8.0]$   & $[2.2,21]$   & $=\tau_-$ \\
    AMAZ         & $[2.0,6.0]$   & $[7.9,15]$   & $=\tau_-$ \\
    \hline
    Other        & $T_I$         & $\tau_-$     & $\tau_+$  \\
    elements     & (°C)          & (years)      & (years)   \\
    \hline
    ASI          & $[4.5,8.7]$   & $[1.9,8.4]$  & $=\tau_-$ \\
    PERM         & $[3.0,6.0]$   & $[5.3,76]$   & $3000$    \\
    \hline
    \end{tabular}
\end{table}

\begin{table}
    \centering
    \caption{\textbf{Sampling ranges for the parameters $\epsilon_{j,i}$ that describe the influence of element $i$ on element $j$.} A positive value indicates a destabilising interaction, a negative value indicates a stabilising interaction, and $\epsilon_{j,i}=0$ indicates that element $i$ has no direct influence on element $i$. }
    \label{tab:interaction_params}
    
    \begin{tabular}{llcccccccc}
    \\
    \hline
    Elements &   & $\epsilon_{\text{GRIS},i}$       & $\epsilon_{\text{WAIS},i}$        & $\epsilon_{\text{EASB},i}$        & $\epsilon_{\text{EAIS},i}$ & $\epsilon_{\text{AMOC},i}$                & $\epsilon_{\text{AMAZ},i}$ & $\epsilon_{\text{ASI},i}$                 & $\epsilon_{\text{PERM},i}$ \\
    \hline
    GRIS  &   & $0$                                & $\left[\tfrac{1}{3},\tfrac{2}{3}\right]$ & $\left[\tfrac{1}{3},\tfrac{2}{3}\right]$ & $0$                        & $\left[2,10\right]$            & $0$                        & $0$                                     & $0$                        \\
    WAIS  &   & $\left[0,\tfrac{1}{3}\right]$      & $0$                                & $0$                                & $0$                        & $\left[-\tfrac{20}{3},\tfrac{20}{3}\right]$ & $0$                        & $0$                                     & $0$                        \\
    EASB  &   & $\left[0,\tfrac{1}{3}\right]$      & $0$                                & $0$                                & $0$                        & $0$                                     & $0$                        & $0$                                     & $0$                        \\
    EAIS  &   & $0$                                & $0$                                & $0$                                & $0$                        & $0$                                     & $0$                        & $0$                                     & $0$                        \\
    AMOC  &   & $\left[-1,-\tfrac{2}{3}\right]$    & $\left[0,1\right]$                 & $\left[0,1\right]$                 & $0$                        & $0$                                     & $\left[-\tfrac{2}{3},\tfrac{2}{3}\right]$ & $\left[-\tfrac{2}{3},0\right]$ & $0$                        \\
    AMAZ  &   & $0$                                & $0$                                & $0$                                & $0$                        & $0$                                     & $0$                        & $0$                                     & $0$                        \\
    ASI   &   & $\left[0,\tfrac{1}{3}\right]$      & $0$                                & $0$                                & $0$                        & $\left[0,\tfrac{2}{3}\right]$            & $0$                        & $0$                                     & $\left[0,\tfrac{1}{3}\right]$ \\
    PERM  &   & $0$                                & $0$                                & $0$                                & $0$                        & $\left[0,1\right]$                       & $0$                        & $0$                                     & $0$                        \\
    \hline
    \end{tabular}
\end{table}

\begin{table}
    \centering
    \caption{\textbf{Fraction of model simulations exhibiting tipping of the Greenland Ice Sheet (GRIS), the West Antarctic Ice Sheet (WAIS), and the East Antarctic Subglacial Basins (EASB), either individually or in combination.} Fractions are given for different emission scenarios as percentages. For each combination of tipped elements among GRIS, WAIS and EASB, we give the approximate global mean sea level rise in the year 100{,}000~CE.}
    \label{tab:scenario_modes}

    \begin{tabular}{lcccccccc}
    \\
    \hline
    Scenario &
    \multicolumn{1}{c}{None} &
    \multicolumn{1}{c}{WAIS} &
    \multicolumn{1}{c}{GRIS} &
    \multicolumn{1}{c}{\shortstack{GRIS\\WAIS}} &
    \multicolumn{1}{c}{EASB} &
    \multicolumn{1}{c}{\shortstack{WAIS\\EASB}} &
    \multicolumn{1}{c}{\shortstack{GRIS\\EASB}} &
    \multicolumn{1}{c}{\shortstack{GRIS\\WAIS\\EASB}} \\
    & $\sim$2 m & $\sim$7 m & $\sim$11 m & $\sim$15 m & $\sim$21 m & $\sim$25 m & $\sim$28 m & $\sim$32 m \\
    \hline
    50~PgC   & 99.60 & 0.20  & 0.00 & 0.00 & 0.10 & 0.10 & 0.00 & 0.00 \\
    300~PgC  & 96.50 & 2.20  & 0.30 & 0.10 & 0.50 & 0.40 & 0.00 & 0.00 \\
    1200~PgC & 46.20 & 23.70 & 3.40 & 9.10 & 4.80 & 9.50 & 0.50 & 2.80 \\
    3800~PgC & 0.00  & 3.20  & 0.00 & 0.70 & 0.00 & 32.60 & 0.00 & 63.50 \\
    5000~PgC & 0.00  & 0.00  & 0.00 & 0.00 & 0.00 & 12.60 & 0.00 & 87.40 \\
    \hline
    \end{tabular}
\end{table}

\begin{table}
    \centering
    \caption{\textbf{Parameter values for SURFER's sea level rise component.}}
    \label{tab:slrparams_1}
    
    \begin{tabular}{lcr}
    \\
    \hline
    Parameter & Associated quantity & Value \\
    \hline
     $h_{\mathrm{U}}$ & upper layer depth & $150$~m\\
 $h_{\mathrm{I}}$ & intermediate layer depth & $500$~m\\
 $h_{\mathrm{D}}$ & deep layer depth & $3150$~m\\
     $\alpha_{\mathrm{U}}$ & thermal expansion coefficient for layer U & $2.20 \times 10^{-4}~\mathrm{K}^{-1}$  \\
     $\alpha_{\mathrm{I}}$ & thermal expansion coefficient for layer I & $1.61 \times 10^{-4}~\mathrm{K}^{-1}$  \\
     $\alpha_{\mathrm{D}}$ & thermal expansion coefficient for layer D & $1.39 \times 10^{-4}~\mathrm{K}^{-1}$  \\
       $S_{\mathrm{gl\:pot}}$    & sea level rise potential from mountain glaciers & $0.5$~m \\
     $\xi$    &  sensitivity coefficient for glacier parametrisation & $2$~°C \\
     $\tau_{\mathrm{gl}}$ & timescale for glacier melt & $200$~yr \\
     $S^{\text{pot}}_{\text{GRIS}}$ & sea level rise potential from Greenland ice-sheet & $7$~m \\
     $S^{\text{pot}}_{\text{WAIS}}$ & sea level rise potential from West Antarctic ice-sheet & $4$~m \\
     $S^{\text{pot}}_{\text{EASB}}$ & sea level rise potential from East Antarctic sublgacial basins & $19$~m \\
     $S^{\text{pot}}_{\text{EAIS}}$ & sea level rise potential from East Antarctic ice-sheet & $34$~m    \\
     \hline
    \end{tabular}
\end{table}


\clearpage 



\end{document}